\documentclass[singlecolumn,amsmath,amssymb,pre,a4paper]{revtex4}
\usepackage{txfonts} 
\usepackage{amsfonts} 
\usepackage{amsmath}
\usepackage{graphicx} 
\usepackage{overpic} 
\usepackage{color}

\newcommand{\be}{\begin{eqnarray}} 
\newcommand{\ee}{\end{eqnarray}}
 
\newcommand{\D}{\mathrm{d}}

\begin{document}

\title{Mean-field theory of random close packings of axisymmetric particles}

\author{Adrian Baule$^{1,2}$, Romain Mari$^{1}$, Lin Bo$^{1}$, Louis Portal$^{1}$
\& Hern\'an A. Makse$^{1}\footnote{Correspondence to: hmakse@lev.ccny.cuny.edu}$}

\affiliation{ $^1$Levich Institute and Physics Department, City
  College of New York, New York, New York 10031, USA \\ $^2$School of
  Mathematical Sciences, Queen Mary University of London, London E1
  4NS, UK}

\date{\today}

\begin{abstract}

Finding the optimal random packing of non-spherical particles is an
  open problem with great significance in a broad range of scientific
  and engineering fields. So far, this search has been performed only
  empirically on a case-by-case basis, in particular, for shapes like
  dimers, spherocylinders and ellipsoids of revolution. Here, we
  present a mean-field formalism to estimate the packing density of axisymmetric non-spherical particles. We derive an analytic
  continuation from the sphere that provides a phase diagram
  predicting that, for the same coordination number, the density of
  monodisperse random packings follows the sequence of increasing packing fractions: spheres $<$ oblate ellipsoids $<$ prolate
  ellipsoids $<$ dimers $<$ spherocylinders. We find the maximal
  packing densities of 73.1\% for spherocylinders and 70.7\% for
  dimers, in good agreement with the largest densities found in
  simulations. Moreover, we find a packing density of 73.6\% for lens-shaped particles, representing the densest random packing of
  the axisymmetric objects studied so far.  

\end{abstract}


\maketitle

Understanding the properties of assemblies of particles from the
anisotropy of their building blocks is a central challenge in
materials science \cite{Glotzer07,Damasceno12,Ni12}. In particular,
the shape that leads to the densest random packing has been
systematically sought empirically
\cite{Williams03,Abreu03,Donev04,Man05,Jia07,Bargiel08,Wouterse09,HajiAkbari09,Faure09,Jaoshvili10,Lu10,Kyrylyuk11,Jiao11,Zhao12},
since it is expected to constitute a superior glass forming material
\cite{Glotzer07}. Despite the significance of random packings of
anisotropic particles in a range of fields like self-assembly of
nanoparticles, liquid crystals, glasses, and granular processing
\cite{Torquato10}, there is yet no theoretical framework to estimate
their packing density. Thus, random packings of anisotropic particles
are typically investigated on a case-by-case basis using computer
simulations, which have shown, e.g., that elongated shapes like
prolate ellipsoids and spherocylinders can pack considerably denser
than the random-close packing (RCP) fraction of spheres at $\phi_{\rm
  RCP}\approx 0.64$. These shapes exhibit a maximum in the packing
fraction for aspect ratios (length/width) close to the sphere
\cite{Williams03,Abreu03,Donev04}.

Table~\ref{table1} summarizes the empirical findings for maximal
densities and highlights a further caveat of simulation and
experimental studies: The protocol dependence of the final
close-packed (or jammed) state leading to a large variance of the
maximal packing fractions found for the same shape. This observation
can be explained using the picture of a rugged energy landscape from
theories of the glass phase \cite{Parisi10}. Different algorithms get
stuck in different metastable basins of the energy landscape, reaching
different final packing states.

Here, we present a mean-field approach
to systematically study the packing fraction of a class of anisotropic
shapes with rotational symmetry, which can therefore guide further
empirical studies. Explicit results are obtained for axisymmetric
particles like dimers, spherocylinders and lens-shaped particles and
we discuss generalizations to other shapes like tetrahedra, cubes and
irregular polyhedra. Furthermore, we derive an analytic continuation
of the spherical RCP which provides a phase diagram for these and
other anisotropic particles like oblate and prolate ellipsoids. We first define the Voronoi volume of a non-spherical particle on
which our calculation is based, and show that it can be calculated
analytically for many different shapes by a decomposition of the shape
into overlapping and intersecting spheres, which we organize into
interactions between points, lines and anti-points. We then develop a
statistical mean-field theory of the Voronoi volume to treat the
particle correlations in the packing. This geometric mean-field
approach is complemented by a quantitative estimation of the variation
of the average contact number with the particle aspect ratio. The
predicted packing density is interpreted as an upper bound of the
empirically obtained packings.

\section*{Results}

\subsection*{Voronoi boundary between non-spherical objects}

We consider rotationally symmetric objects for which the aspect ratio
$\alpha$ is defined as length/width, where the length is measured
along the symmetry axis. In the following, we focus on the region $0<\alpha <2$, where the largest densities are found \cite{Zhao12}. Our description of
packings relies on a suitable tessellation of space into
non-overlapping volumes \cite{Song08}. We use the standard Voronoi
convention \cite{Aurenhammer91,Okabe}, where one associates with each
particle the fraction of space that is closer to this particle than to
any other one. This defines the Voronoi volume $W_i$ of a particle
$i$, which depends on the configurations
$\mathbf{x}=(\mathbf{r},\mathbf{\hat{t}})$ of all particles (including
position $\mathbf{r}$ and orientation $\mathbf{\hat{t}}$). The total
volume $V$ occupied by $N$ particles is $ V=\sum_{i=1}^N
W_i(\{\mathbf{x}_1,...,\mathbf{x}_N\})$, and the packing fraction of
monodisperse particles of volume $V_\alpha$ and aspect ratio $\alpha$
follows as $\phi=N V_\alpha/V$. In order to determine $W_i$ one has to
know the Voronoi boundary (VB) between two particles $i$ and $j$,
which is the hypersurface that contains all points equidistant to both
particles (Fig.~\ref{Fig_coord} for
spherocylinders). The VB of the volume $W_i$ along $\mathbf{\hat{c}}$,
denoted by $l_i(\mathbf{\hat{c}})$, is the minimal one in this
direction among all possible VBs of each particle $j$ in the
packing. It is formally obtained by the global minimization
\cite{Song08}: \be
\label{globmin}
l_i(\mathbf{\hat{c}})= \min_{j:s>0}\;s(\mathbf{r}_j,\mathbf{\hat{t}}_j,\mathbf{\hat{c}}), \ee where
$s(\mathbf{r}_j,\mathbf{\hat{t}}_j,\mathbf{\hat{c}})$ denotes the VB along $\mathbf{\hat{c}}$
between particles $i$ and $j$ with relative position $\mathbf{r}_j$ and
orientation $\mathbf{\hat{t}}_j$ (Fig.~\ref{Fig_coord}). The Voronoi volume
follows then exactly as the orientational integral, \be
\label{methods_wi}
W_i=\frac{1}{3}\oint\D \mathbf{\hat{c}} \,l_i(\mathbf{\hat{c}})^3.
\ee

The VB between two equal spheres is identical to the VB between two points and is a flat plane
perpendicular to the separation vector
(Fig.~\ref{Fig_shapes}a) \cite{Song08}. Finding the VB for more complicated shapes
is a challenging problem in computational geometry, which is typically
only solved numerically \cite{Boissonnat06}. We approach this problem
analytically by considering a decomposition of the non-spherical shape
into overlapping spheres. The VB is then determined as follows: Every
segment of the VB arises due to the Voronoi interaction between a
particular sphere on each of the two particles reducing the problem to
identifying the correct spheres that interact. This identification
follows an exact algorithm for a large class of shapes obtained by the
union and intersection of spheres, which can be translated into an
analytical expression of the VB as outlined in
Fig.~\ref{Fig_algorithms} for dimers, spherocylinders and lens-shaped
particles.

For instance, a dimer is the union of a pair of spheres (Fig.~\ref{Fig_shapes}b). The
  dimers VB is thus a composition of maximal four different surfaces
  depending on the relative orientation of the dimers defined by four
  points at the centre of each sphere
  (Fig.~\ref{Fig_algorithms}a). The extension to trimers is
  straightforward (Fig.~\ref{Fig_shapes}c). Likewise, $n$ overlapping
  spheres lead to compositions of $n$ surfaces. A spherocylinder is a
  dense overlap of spheres of equal radii and the VB interaction is
  identical to that between four points and two lines
  (Fig.~\ref{Fig_shapes}d). The interactions then simplify into
  line-line, line-point, and point-point interactions, which generally
  lead to a curved VB for non-parallel orientations
  (Fig.~\ref{Fig_algorithms}b). 

The Voronoi decomposition used for dimers and spherocylinders can be generalized to arbitrary shapes by using a dense filling
of spheres with unequal radii \cite{Phillips12}. However, even if it
is still algorithmically well defined, this procedure may become
practically tedious for dense unions of polydisperse spheres. Alternatively one can apply
specialized algorithms to compute numerical VBs between curved line segments
\cite{Hoff99}. Here, we propose an analytically tractable approach: Convex shapes can be approximated by intersections of a finite number of spheres. An
oblate ellipsoid, e.g., is well approximated by a lens-shaped
particle, which consists of the intersection of two spheres; an
intersection of four spheres is close to a tetrahedra, and six spheres
can approximate a cube. This is illustrated in
Fig.~\ref{Fig_shapes}e--h, and the corresponding algorithms outlined in Fig.~\ref{Fig_algorithms}c. The main insight is that the effective Voronoi interaction of these
shapes is governed by a symmetry: Points map to ``anti-points'' (since
the interactions between spheres is inverted;
Fig.~\ref{Fig_algorithms}c). The VB of ellipsoid-like objects arises
from the interaction between four anti-points and four points in two
dimensions (Fig.~\ref{Fig_algorithms}c) or lines in three dimensions,
and thus falls into the same class as spherocylinders. For cubes the
effective interaction is that of twelve lines, eight points and six
anti-points (Fig.~\ref{Fig_shapes}g). Analytic expressions of the VB
for dimers and spherocylinders are calculated in the Supplementary Methods.
  
  \subsection*{A statistical theory for Voronoi volume
  fluctuations}
  
We turn the above formalism into a mean-field theory to calculate the
volume fraction of a packing of monodisperse non-spherical objects. In order to take into
account multi-particle correlations in the packing, we use a
statistical mechanics treatment where the overall volume is expressed
in terms of the average Voronoi volume $\overline{W}(z)$:
$V=N\overline{W}(z)$ \cite{Song08} characterized by the average
coordination number $z$, which denotes the mean number of contacting
neighbours in the packing. This approach is motivated by the
observation that, as $N\to\infty$, packings exhibit reproducible phase
behaviour, which is characterized by only few observables such as
$\phi$ and $z$ \cite{Makse04}. Our statistical mechanics framework is
based on the Edwards ensemble approach, which considers the
volume as a Hamiltonian of the system and attempts to find the minimum
volume \cite{Edwards89}. Here, $\overline{W}$ is
given as the ensemble average of $W_i$ over all particles in the
packing: $\overline{W}=\left< W_i\right>_i$. We obtain therefore from
Eq.~(\ref{methods_wi}): \be \overline{W}&=&\left<\frac{1}{3}\oint\D
\mathbf{\hat{c}}\, l_i(\mathbf{\hat{c}})^3\right>_i=\frac{1}{3}\oint\D
\mathbf{\hat{c}}\, \left<l_i(\mathbf{\hat{c}})^3\right>_i =\nonumber\\
\label{methods_avw}
&=&\frac{1}{3}\oint\D \mathbf{\hat{c}}\int_{c^*(\mathbf{\hat{c}})}^\infty \D c\,c^3
p(\mathbf{c}).  \ee In the last step we have introduced the probability
density $p(\mathbf{c})$ which contains the probability to find the VB at
$c$ in the direction $\mathbf{\hat{c}}$. The lower integration limit
$c^*(\mathbf{\hat{c}})$ is the minimal value of the boundary along $\mathbf{\hat{c}}$,
which corresponds to the hard core boundary of the particle in that
direction. We introduce the cumulative distribution function (CDF)
$P(\mathbf{c})$ via the usual definition $p(\mathbf{c})=-\frac{\D}{\D
  c}P(\mathbf{c})$. Substituting the CDF in Eq.~(\ref{methods_avw}) and
performing an integration by parts leads to the volume integral \be
\label{integral}
\overline{W}(z)=\int\D\mathbf{c}\,P(\mathbf{c},z), \ee where we
indicate the dependence on $z$. In a geometric picture \cite{Song08},
$P(\mathbf{c},z)$ is interpreted as the probability that $N-1$
particles are outside a volume $\Omega$ centered at $\mathbf{c}$ (see
Fig.~\ref{Fig_volumes}), since otherwise they would contribute a
shorter VB. This leads to the definition \be
\label{methods_vex}
\Omega(\mathbf{c},\mathbf{\hat{t}})=\int \D
\mathbf{r}\,\Theta(c-s(\mathbf{r},\mathbf{\hat{t}},\mathbf{\hat{c}}))\Theta(s(\mathbf{r},\mathbf{\hat{t}},\mathbf{\hat{c}})),
\ee where $\Theta(x)$ denotes the usual Heavyside step. We refer to
$\Omega$ as the Voronoi excluded volume, which extends the standard
concept of the hard-core excluded volume $V_{\rm ex}$ considered by
Onsager in his theory of elongated equilibrium rods \cite{Onsager49} (Fig.~\ref{Fig_volumes}).

The dependence of $P(\mathbf{c},z)$ on $\Omega$ has been treated at a
mean-field level in \cite{Song08} and has been derived from a theory
of correlations using liquid state theory in \cite{Jin10b} for
high-dimensional sphere packings. In both cases it provides a
Boltzmann-like exponential form $P(\mathbf{c},z)\propto
\exp\left\{-\int_{\Omega(\mathbf{c})}\D\mathbf{r}\,\rho(\mathbf{r},z)\right\}$
in the limit $N\to\infty$, where $\rho(\mathbf{r},z)$ is the density of
spheres at $\mathbf{r}$.

The crucial step is to generalize this result to anisotropic
particles. Following Onsager \cite{Onsager49}, we treat particles of
different orientations as belonging to different species. This is the
key assumption to treat orientational correlations within a mean-field
approach. Thus, the problem for non-spherical particles can be mapped
to that of polydisperse spheres for which $P$ factorizes into the
contributions of the different radii \cite{Danisch10}. We thus obtain
the factorized form: \be
\label{supP_cdf}
P(\mathbf{c},z)=\exp\left\{-\int\D\mathbf{\hat{t}}\int_{\Omega(\mathbf{c},\mathbf{\hat{t}})}\D\mathbf{r}\,\rho(\mathbf{r},\mathbf{\hat{t}},z)\right\},
\ee where $\rho(\mathbf{r},\mathbf{\hat{t}},z)$ is the density of
particles with orientation $\mathbf{\hat{t}}$ at $\mathbf{r}$. 

Next, we assume an approximation of this density in terms of contact
and bulk contributions, which is motivated by the connection with the
radial distribution function in spherical theories in both high and
low dimensions \cite{Song08,Jin10b}. The contact contribution relies
on the condition of contact between two particles of a given relative
position $\mathbf{r}$ and orientation $\mathbf{\hat{t}}$, which
defines the contact radius $r^*(\mathbf{\hat{r}},\mathbf{\hat{t}})$:
$r^*$ is the value of $r$ for which the two particles are in contact
without overlap. In the case of equal spheres the contact radius is
simply $r^*(\mathbf{\hat{r}},\mathbf{\hat{t}})=2a$. For non-spherical objects,
$r^*(\mathbf{\hat{r}},\mathbf{\hat{t}})$ depends on the object shape and the relative
orientation (Supplementary Methods). Using $r^*(\mathbf{\hat{r}},\mathbf{\hat{t}})$ we can
separate bulk and contact terms in
$\rho(\mathbf{r},\mathbf{\hat{t}},z)$ as in \cite{Song08,Jin10b}: \be
\label{exactdensity}
\rho(\mathbf{r},\mathbf{\hat{t}},z)=\frac{1}{4\pi}\left[\overline{\rho}\,\Theta(r-r^*(\mathbf{\hat{r}},\mathbf{\hat{t}}))+\sigma(z)\delta(r-r^*(\mathbf{\hat{r}},\mathbf{\hat{t}}))\right].
\ee The prefactor $1/4\pi$ is the density of orientations, which we
assume isotropic. The symbols $\overline{\rho}$ and $\sigma(z)$ stand
for the average free-volume of particles in the bulk and the average
free-surface of particles at contact, respectively, which are discussed further below. The approximation
Eq. (\ref{exactdensity}) corresponds to considering a pair
distribution function as a delta function modeling the contact
particles plus a constant term modeling the particles in the bulk
\cite{Jin10b}, which are thus considered as a uniform structure. These
assumptions are further tested in the Methods section.

Substituting Eq.~(\ref{exactdensity}) into Eq.~(\ref{supP_cdf}) leads
to our final result for the CDF:
\be
\label{cdf}
P(\mathbf{c},z)=\exp\left\{-\overline{\rho}(\overline{W})
\,V^*(\mathbf{c})-\sigma(z)\,S^*(\mathbf{c})\right\}. \ee Here, we
have explicitly written the dependence of $\overline{\rho}$ on $\overline{W}$,
which is important to interpret Eq. (\ref{integral}) as a
self-consistent equation to obtain the volume fraction of the packing. The free volume per particle in the bulk depends
specifically on $\overline{W}(z)$ as $\overline{\rho}=
1/(\overline{W}(z)-V_\alpha)$.

The CDF thus factorizes into two contributions: A contact term:
\be
\label{pc}
P_{\rm C}(\mathbf{c},z)=\exp\left\{-\sigma(z)\,S^*(\mathbf{c})\right\},
\ee
and a bulk term:
\be
\label{pb}
P_{\rm B}(\mathbf{c})=\exp\left\{-\overline{\rho}(\overline{W})\,V^*(\mathbf{c})\right\},
\ee
 such that
\begin{equation}
\label{fact}
P(\mathbf{c},z) = P_{\rm C}(\mathbf{c},z) \times P_{\rm B}(\mathbf{c}).
\end{equation}
The volume $V^*$ is the volume excluded by $\Omega$ for bulk particles
and takes into account the overlap between $\Omega$ and the hard-core
excluded volume $V_{\rm ex}$: $V^*=\left<\Omega-\Omega\cap V_{\rm
  ex}\right>_{\mathbf{\hat{t}}}$, where
$\left<...\right>_{\mathbf{\hat{t}}}$ denotes an orientational
average. Likewise, $S^*$ is the surface excluded by $\Omega$ for
contacting particles: $S^*=\left<\partial V_{\rm
  ex}\cap\Omega\right>_{\mathbf{\hat{t}}}$, where $\partial V_{\rm
  ex}$ denotes the boundary of $V_{\rm ex}$. The volumes $V_{\rm ex}$
and $\Omega$ as well as the resulting $V^*$ and $S^*$ are calculated in the Supplementary Methods and shown in Fig.~\ref{Fig_volumes} for
spherocylinders.

The surface density $\sigma(z)$ is a
measure for the available surface for contacts when the packing is
characterized by an average coordination number $z$. We evaluate this
density by simulating random local configurations of one particle with
$z$ non-overlapping contacting particles and determining the average
available free surface. This surface is given by $S^*(\mathbf{c}_{\rm m})$,
where $c_{\rm m}$ is the minimal contributed VB among the $z$ contacts in
the direction $\mathbf{\hat{c}}$. Averaging over many realizations
with a uniform distribution of orientations and averaging also over
all directions $\mathbf{\hat{c}}$ provides the surface density in the
form, \be
\label{dens2}
\sigma(z)&=&\frac{1}{\left<\left<S^*(\mathbf{c}_{\rm m})\right>\right>_{\mathbf{\hat{c}}}}.
\ee In this way we can only calculate $\sigma(z)$ for integer values
of $z$. For fractional $z$ that are predicted from our evaluation of
degenerate configurations in the next section, we use a linear
interpolation to obtain $\overline{W}(z)$.

Equations~(\ref{integral}) and (\ref{cdf}) lead to a self-consistent
equation for the average Voronoi volume $\overline{W}(z)$ in the form:
$\overline{W}(z) = {\cal F}[\overline{W}(z)]$. Analytic expressions
for $V^*$ and $S^*$ can be derived in the spherical limit in closed
form, where also the self-consistency equation can be solved exactly
\cite{Song08}. For non-spherical shapes we resort to a numerical
integration to obtain $V^*$ and $S^*$. Equation~(\ref{integral}) can then be solved
numerically, which yields $\overline{W}(z)$, and subsequently the
equation of state for the volume fraction versus coordination number,
$\phi(z,\alpha)=V_\alpha/\overline{W}(z)$, in numerical form (denoting explicitly the dependence on $\alpha$).

\subsection*{Variation of the coordination number with aspect ratio}

In this purely geometric theory of the average Voronoi volume, the
packing fraction is given as $\phi(z,\alpha)$, with $z$ and $\alpha$
free parameters, in principle. In practice, $z$ is fixed by the
symmetry properties of the object shape, $z(\alpha)$, and the physical
condition of mechanical stability, requiring force and torque balance
on every particle. Under the assumption of minimal correlations, these
conditions typically motivate the isostatic conjecture based on
Maxwell's counting argument \cite{Alexander98}: $z=2d_{\rm f}$, with $d_{\rm f}$
the number of degrees of freedom, giving $z=6$ for fully symmetric
objects (spheres), $z=10$ for rotationally symmetric shapes like
spherocylinders, dimers and ellipsoids of revolution \cite{Donev04},
and $z=12$ for shapes with three different axis like aspherical
ellipsoids and tetrahedra \cite{Jaoshvili10}. While the isostatic
conjecture is well-satisfied for spheres, packings of non-spherical
objects are in general hypoconstrained with $z<2 d_{\rm f}$, where
$z(\alpha)$ increases smoothly from the spherical value for $\alpha>1$
\cite{Donev04}. The fact that these packings are still in a
mechanically stable state can be understood in terms of the occurrence
of stable degenerate configurations (Fig.~\ref{Fig_spheroz}), which reduce the effective number of degrees of
freedom \cite{Donev07}. However, the observed variation $z(\alpha)$
could not be explained quantitatively so far. Here, we deduce the
relation $z(\alpha)$ by evaluating the probability of finding these
degenerate configurations to provide a prediction of $\phi(\alpha)$ in
close form.

In a degenerate configuration, force balance already implies torque
balance, since the net forces are aligned with the inner axis of the
particle (Figs.~\ref{Fig_spheroz}). This
implies that there is redundancy in the set of force and torque
balance equations for mechanical equilibrium since force and torque
balance equations are not linearly independent. Our
evaluation of these degenerate configurations is based on the assumption that a particle is
always found in an orientation such that the redundancy in the
mechanical equilibrium conditions is maximal. This condition
allows us to associate the number of linearly independent equations
involved in mechanical equilibrium with the set of contact
directions. Averaging over the possible sets of contact directions
then yields the average effective number of degrees of freedom
$\tilde{d}_{\rm f}(\alpha)$, from which the coordination number follows as
$z(\alpha)=2\tilde{d}_{\rm f}(\alpha)$ (Methods).

The results for $z(\alpha)$ are shown in Fig.~\ref{Fig_results}a for
prolate ellipsoids of revolution, spherocylinders, dimers, and
lens-shaped particles. We are able to recover the observed continuous
transition as a function of $\alpha$ from the isostatic coordination
number for spheres, $z=6$ at $\alpha=1$, to the isostatic value $z=10$, for aspect ratios above
$\approx 1.5$. The trend compares well to known data for ellipsoids
\cite{Donev04} and spherocylinders \cite{Zhao12,Wouterse09}. In
particular, our approach explains the decrease of $z$ for higher
aspect ratios observed in simulations of spherocylinders
\cite{Zhao12,Wouterse09}: For large $\alpha$, the most probable case
is to have contacts only on the cylindrical part of the particle, so
that all normal forces are coplanar reducing the effective number
of degrees of freedom by one. Consequently, $z \to 8$ as $\alpha \to
\infty$, as we obtain in Fig.~\ref{Fig_results}a. This decrease is
specific to spherocylinders, and not observed for dimers or
ellipsoids, since the normal forces are not coplanar.

\subsection*{Phase diagram of non-spherical particles}

Our calculation leads to a close theoretical prediction for the
packing density $\phi(\alpha)=\phi(z(\alpha),\alpha)$ which does not
contain any adjustable parameters. Figure~\ref{Fig_results}b shows the
prediction for dimers, spherocylinders, and lens-shaped particles. For
spherocylinders, results in the literature on $\phi(\alpha)$ vary
greatly (Table~\ref{table1}), but all show a peak at around $\alpha\approx
1.3-1.5$, which is captured by our formalism. We predict the maximum
density of spherocylinders at $\alpha = 1.3$ with a density $\phi_{\rm
  max} = 0.731$ and that of dimers at $\alpha= 1.3$ with $\phi_{\rm
  max} =0.707$. We have also calculated the packing fraction of the
lens-shaped particles of Fig.~\ref{Fig_algorithms}c, which yields
$\phi_{\rm max}=0.736$ for $\alpha=0.8$. This shape represents the
densest random packing of an axisymmetric shape known so far.

We further investigate packings of non-spherical objects in the
$z$-$\phi$ representation. This change in perspective allows us to
characterize packings of differently shaped objects in a phase
diagram. By plotting $z(\alpha)$ against $\phi(\alpha)$ parametrically
as a function of $\alpha$, we obtain a phase diagram for jammed
anisotropic particles in the $z$-$\phi$ plane
(Fig.~\ref{Fig_results}c). In the same diagram, we also plot the
equation of state obtained with the present theory in the case of
spheres in \cite{Song08}: $\phi_{\rm sph}(z)=z/(z+2\sqrt{3})$, which
is valid between the two isostatic limits of frictionless spheres
$z=6$ and infinite frictional spheres at $z=4$. Surprisingly, we find
that both dimer and spherocylinder packings follow an analytical
continuation of these spherical packings. This result highlights that
the spherical random branch can be continued smoothly beyond the RCP
in the $z$-$\phi$ plane.

The analytical continuation of RCP is derived by solving the
self-consistent Eq.~(\ref{integral}) close to the spherical limit (Supplementary Methods): \be
\label{continuation}
\phi(z)=\left(1+\omega_1\frac{1+g_1(\omega_1)\left(\frac{z}{\bar{z}}-1\right)\frac{M_{\rm b}}{M_z}}{\left[\frac{z}{\bar{z}}-g_2(\omega_1)\left(\frac{z}{\bar{z}}-1\right)\frac{M_{\rm b}}{M_z}\right]\left[1+\left(\frac{z}{\bar{z}}-1\right)\frac{M_{\rm v}}{M_z}\right]}\right)^{-1}.
\ee Here, $\omega_1 = 1/\sqrt{3}$ denotes the spherical free volume at
RCP defined as $\omega_1=1/\phi_{\rm sph} - 1$ evaluated at $z=6$ as
calculated in \cite{Song08}, $\bar{z}=6$ is the spherical isostatic
value, and the functions $g_{1,2}$ can be expressed in terms of
exponential integrals. The dependence of Eq.~(\ref{continuation}) on
the object shape is entirely contained in the geometrical parameters
$M_{\rm b}$, $M_{\rm v}$, and $M_z$: $M_{\rm b}$ and $M_{\rm v}$ quantify the first order
deviation from the sphere at $\alpha=1$ of the object's hard-core
boundary and its volume, respectively, while $M_z$ measures the first
order change in the coordination number upon deformation of the
sphere. The resulting continuations $z(\phi)$ obtained by inverting
Eq.~(\ref{continuation}) for different object shapes are plotted in
the inset of Fig.~\ref{Fig_results}c.

For the smooth shapes considered, we find generally that denser
packing states are reached for higher coordination numbers. For a
given value of $z$, spherocylinders achieve the densest packing,
followed by dimers, prolate ellipsoids, and oblate ellipsoids, as seen
in the inset of Fig.~\ref{Fig_results}c. We observe that the densest packing states for dimers and
spherocylinders found in simulations lie
almost exactly on the continuation, while the one of the ellipsoids
deviate considerably.

\subsection*{Comparison with empirical data}

Table~\ref{table1} indicates that there
is a finite range of densities for random jammed packings according to the particular experimental or numerical
protocol used (denoted as a J-line in the case of jammed spheres
\cite{Parisi10,Skoge06}). On the other hand, our mean-field theory
predicts a single density value and Fig.~\ref{Fig_results} indicates
that our predictions are an upper bound of the empirical
results. We interpret these results in terms of current views of
the jamming problem developed in the limiting case of spheres, where
the question of protocol-dependency of packings has been
systematically investigated.

Random close packings can be considered as infinite-pressure limits of
metastable glass states, which was shown theoretically in
\cite{Krzakala2007,Mari09,Biazzo09,Parisi10} and confirmed in computer
simulations in \cite{Hermes10}. Indeed, there exist a range of packing
fractions named as $[\phi_{\rm th}, \phi_{\rm GCP}]$ following the
notation of mean-field Replica Theory (RT) \cite{Parisi10}. Here,
$\phi_{\rm GCP}$ stands for the density of the ideal glass close
packing and is the maximum density of disordered packings, while
$\phi_{\rm th}$ is the infinite-pressure limit of the least dense
metastable states. In RT, the states $[\phi_{\rm th}, \phi_{\rm GCP}]$
are all isostatic.

 From the point of view of simulations, the well-known
 Lubachevsky-Stillinger (LS) protocol \cite{Skoge06} provides this
 range of packings for different compression rates. The densities
   $[\phi_{\rm th}, \phi_{\rm GCP}]$ are achieved by the corresponding
   compression rates (from large to small) $[\gamma_{\rm th},
   \gamma_{\rm GCP} \to 0]$.  Compression rates
   larger than $\gamma_{\rm th}$ all end to $\phi_{\rm th}$. The
   threshold value $\gamma_{\rm th}$ corresponds to the relaxation
   time $1/\gamma_{\rm th}$ of the least dense metastable
   glass states.  The denser states at GCP are unreachable by
 experimental or numerically generated packings, as it
   requires to equilibrate the system in the ideal glass phase, a
   region where the relaxation time is infinite. In general, large
 compression rates lead to lower packing fractions. This picture was
 investigated for sphere packings in \cite{Skoge06,Chaudhuri10} and it
 is particularly valid for high dimensional systems where
 crystallization is avoided \cite{Parisi10}.

Random close packings are also known to display sharp structural
changes \cite{Anikeenko07,Radin08,Jin10,Klumov11,Kapfer12} signalling
the onset of crystallization at a freezing point $\phi_c$
\cite{Torquato10}. All the (maximally random) jammed states along the
segment $[\phi_{\rm th}, \phi_{\rm GCP}]$ can be made denser at the
cost of introducing some partial crystalline order.  Support for a
order/disorder transition at $\phi_c$ is also obtained from the
increase of polytetrahedral substructures up to RCP and its consequent
decrease upon crystallization \cite{Anikeenko08}. In terms of protocol
preparation like the LS algorithm, there exists a typical time scale
$t_c$ corresponding to crystallization. Crystallization appears in LS
\cite{Parisi10,Torquato10,Jin10} if the compression rate is smaller
than $\gamma_c =1/t_c$, around the freezing packing
fraction~\cite{Cavagna2009}.  A possible path to avoid crystallization
and obtain RCP in the segment $[\phi_{\rm th}, \phi_{\rm GCP}]$ is to
equilibrate with $\gamma > \gamma_c$ to pass the freezing point, and
eventually setting the compression rate in the range $[\gamma_{\rm
    th}, \gamma_{\rm GCP} \to 0]$ to achieve higher volume fraction.

Since the present statistical mechanics framework is based on the
Edwards ensemble approach \cite{Edwards89}, our prediction of the
packing density $\phi_{\rm Edw}$ corresponds to the ensemble average
over the configuration space of random states at a fixed coordination
number.  Since the volume plays the role of the Hamiltonian, the
energy minimization in equilibrium statistical mechanics is replaced
in our formalism by a volume minimization: The highest volume fraction
for a given disordered system is achieved in the limit of zero
compactivity. Therefore, the present framework provides a mean-field
estimation of such a maximal volume fraction (minimum volume) of
random packings with no crystallization. As we perform an ensemble
average over all packings at a fix coordination number, the obtained
volume fraction $\phi_{\rm Edw}$ corresponds to the one with the
largest entropy (called largest complexity in RT) along $[\phi_{\rm
    th}, \phi_{\rm GCP}]$. This point needs not to be $\phi_{\rm th}$,
and in general it is a larger volume fraction. Thus, $\phi_{\rm
  th}<\phi_{\rm Edw}< \phi_{\rm GCP}$.

The above discussion can be translated to the present case of
non-spherical particles. In this case, unfortunately, there is no
detailed study of the protocol dependent packing density as done by
\cite{Skoge06,Chaudhuri10,Parisi10} for spheres. However, the survey
of the available simulated data obtained by different groups (Table 1
and Fig.~\ref{Fig_results}b, c) can be interpreted analogously as for
spheres. In the case of spherocylinders, packings have been obtained
in the range [0.653, 0.722] (these minimum and maximum values have
been obtained in \cite{Abreu03} and \cite{Zhao12}, respectively, see
Table~\ref{table1}). Our predicted density is 0.731, representing an
upper bound to the simulated results. In the case of dimers, there are
two simulations giving a density of 0.697 (Schreck \& O'Hern 2011, personal communication) and 0.703
\cite{Faure09}, which are both smaller than and very close to our
prediction 0.707. Thus, our prediction is interpreted as the upper
limit in the range of packings observed with numerical
algorithms. Under this scenario, which is consistent with analogous 3d
spherical results, packings may exist in the region $[\phi_{\rm th},
  \phi_{\rm Edw}]$, and our theory is a mean-field estimation of
$\phi_{\rm Edw}$. This region is very small for spheres but the above evidence
indicates that non-spherical particles may pack randomly in a broader
range of volumes. The present framework estimates the upper bound for
such a range.

\section*{Discussion}

We would like to stress that our analytic continuation is non-rigorous
and appears as the solution of our mean-field theory for first-order
deviations in $\alpha$ from the sphere using suitable
approximations. The shapes of dimers, spherocylinders, ellipsoids are
then all shown to increase the density of the random packing to
first-order. In the case of regular (crystal) packings, recent
  mathematically rigorous work has shown in fact that for axisymmetric
  particles any small deformation from the sphere will lead to an
  increase in the optimal packing fraction of the crystal
  \cite{Kallus12}. This appears only in 3d and is
related to Ulam's conjecture stating that the sphere is the worst case
scenario for ordered packings in 3d \cite{Gardner}. A full
mathematical proof of this conjecture is still outstanding, but so far
all computer simulations verify the conjecture. In particular, recent
advances in simulation techniques allow to generate crystal packings
of a large variety of convex and non-convex objects in an efficient
manner \cite{Graaf11,Graaf12}. The extensive study of
Ref.~\cite{Graaf11} has extended the verification of Ulam's conjecture
to the first 8 regular prisms and antiprisms, the 92 Johnson solids,
and the 13 Catalan solids. The verification for regular $n$-prisms and
$n$-antiprisms can be extended to arbitrary $n$ using this method,
providing an exhaustive empirical verification of the conjecture for
these regular shapes. We remark that a random analogue of Ulam's
packing conjecture has been proposed and verified for the Platonic
solids (apart from the cube) in simulations \cite{Jiao11}. The results
presented here support the random version of Ulam's conjecture and
might help in investigating this conjecture further from a theoretical
point of view.

We believe that our decomposition of various shapes into intersections
and overlaps of spheres will be a useful starting point for a
systematic investigation of this issue. Our approach can be
systematically continued beyond the axisymmetric shapes considered
here. For instance, in Fig.~\ref{Fig_shapes}e--h, we have 2,3,6,n
anti-points to describe ellipsoids and polyhedra of increasingly
varying complexity. The challenge would be to implement our algorithm
to calculate the resulting Voronoi excluded volumes that appear in our
mean-field theory. For this, one might also consider a fully numerical
evaluation using, e.g., graphics hardware
\cite{Hoff99}.

\section*{Methods}

\subsection*{Quantitative method to calculate $z(\alpha)$}

Mathematically, we can write the local mechanical equilibrium on a
generic non-spherical frictionless particle having $k$ contacts
defined by their location $\mathbf{r}_j$, normal $\mathbf{\hat{n}}_j$,
and force $f_j\mathbf{\hat{n}}_j$, as:
\begin{equation}\label{eq:mech_eq}
  \left(\begin{array}{ccc}
    \mathbf{\hat{n}}_1 & \dots & \mathbf{\hat{n}}_k \\
    \mathbf{r}_1 \times \mathbf{\hat{n}}_1 & \dots & \mathbf{r}_k \times \mathbf{\hat{n}}_k
   \end{array} \right) 
  \left(\begin{array}{c}
    f_1\\
    \vdots\\
    f_k
  \end{array} \right)  
 \equiv \underline{\underline{N}}\, \underline{f}=0, 
\end{equation}
where $\underline{\underline{N}}$ is a $d_{\rm f} \times k$ matrix.  A local
degenerate configuration has a matrix $\underline{\underline{N}}$ such
that ${\rm rank}(\underline{\underline{N}}) < \min(d_{\rm f}, k)$.  We base
our evaluation on two assumptions:
{\it (i)} Contact directions around a particle in the packing are
uncorrelated, and {\it (ii)} Given one set of contact directions, a
particle $i$ is found in an orientation $\mathbf{\hat{t}}_i$ such that
the redundancy in the mechanical equilibrium conditions is maximal,
i.e., ${\rm rank}(\underline{\underline{N}}_{\, \mathbf{\hat{t}}_i})$
is a minimum. Note that $\underline{\underline{N}}_{\,
  \mathbf{\hat{t}}_i}$ depends on $\mathbf{\hat{t}}_i$, as only the
absolute direction of contact points are chosen, and thus rotating
particle $i$ affects the direction and normal of these contacts with
respect to particle $i$. This situation is described in
Fig.~\ref{Fig_spheroz}c, which includes a two-dimensional sketch of a
three dimensional degenerate configuration that we observe often in
our procedure. In this case the rank is reduced by one unit, and the
probability of occurrence of such a situation is large at small aspect
ratio, as it just requires that there is no contact on the cylindrical
part of the inner particle.

Within our assumptions, we explore the space of possible contact
directions for one particle, given a local contact number $k$, and
aspect ratio $\alpha$. We then extract the average effective
number of degrees of freedom $\tilde{d}_{\rm f}(\alpha, k)$, which is the
average over the contact directions of the minimal value of ${\rm
  rank}(\underline{\underline{N}}_{\mathbf{\hat{t}}})$: $\tilde{d}_{\rm f}(\alpha,
k)= \left< \min_{\mathbf{\hat{t}}} \left({\rm rank}(
\underline{\underline{N}}_{\, \mathbf{\hat{t}}}) \right) \right>_{\{\mathbf{r}_1,
  \dots, \mathbf{r}_k \}}$, where $\left<\,...\,\right>_{\{\mathbf{r}_1,
  \cdots, \mathbf{r}_k \}} = \mathcal{N}^{-1} \int_J...  \; \D \mathbf{r}_1
\cdots \D \mathbf{r}_k $ denotes the average over contact
directions. This average is limited to a subset $J$ of all possible
$\{\mathbf{r}_1, \dots, \mathbf{r}_k \}$ such that mechanical equilibrium
(Eq.~\ref{eq:mech_eq}) is possible with positive forces, as expected
for a packing of hard particles. This corresponds geometrically to
sets $\{\mathbf{r}_1, \dots, \mathbf{r}_k \}$ which do not leave a
hemisphere free on the unit sphere. Finally, the normalization
$\mathcal{N}$ is the volume of $J$. For a packing with a coordination
number distribution $Q_z(k)$, with average $z$, the effective $d_{\rm f}$ is: $\tilde{d}_{\rm f}(\alpha)=\sum_k Q_z(k)
\tilde{d}_{\rm f}(\alpha, k)$, and the average $z$ follows as $z(\alpha)=2\tilde{d}_{\rm f}(\alpha)$. In our evaluation, we use a
Gaussian distribution for $Q_z(k)$, with variance $1.2$ and average
$z$, consistent with simulations \cite{Wang11}. Overall, $z(\alpha)$
is thus the solution of the following self-consistent relation:
 \begin{equation}
   z(\alpha) = 2 \sum_k Q_z(k) \left< \min_{\mathbf{\hat{t}}} \left({\rm rank}( \underline{\underline{N}}_{\, \mathbf{\hat{t}}})  \right) \right>_{\{\mathbf{r}_1, \dots, \mathbf{r}_k \}}.
 \end{equation}

The way we look for the orientation $\mathbf{\hat{t}}$ on the unit sphere
showing the lowest rank is simply by sampling it randomly with a
uniform distribution ($10^6$ samples). The computation of the rank is done via a standard
Singular Value Decomposition of $\underline{\underline{N}}_{\,
  \mathbf{\hat{t}}}$, which is here numerically accurate for $\alpha\ge 1.05$.

\subsection*{Test of the approximations of the theory}

We perform a comprehensive
test of the different approximations of the theory using computer
simulations of spherocylinder packings (Supplementary Note~1). From the generated configurations at the jamming point we obtain the CDF $P(\mathbf{c},z)$, where $z$ is also an
observable of the simulation determined by the jamming condition. $P(\mathbf{c},z)$ contains the probability that the boundary of the Voronoi volume in the direction $\mathbf{\hat{c}}$ is found at
a value larger than $c$ and is determined as follows. We select an orientation $\mathbf{\hat{c}}$ relative to the
orientation $\mathbf{\hat{t}}_i$ of a chosen reference particle $i$. A large
number of particles in the packing contribute a VB along $\mathbf{\hat{c}}$
with particle $i$. We determine all these different VBs denoted by
$s(\mathbf{r}_j,\mathbf{\hat{t}}_j,\mathbf{\hat{c}})$. The boundary of the Voronoi volume
in the direction $\mathbf{\hat{c}}$ is the minimum $c_{m}$ of all positive VBs:
\be c_{\rm m}=\min_{j:s>0}s(\mathbf{r}_j,\mathbf{\hat{t}}_j,\mathbf{\hat{c}}), \ee where
$\mathbf{r}_j$ and $\mathbf{\hat{t}}_j$ are the relative position and orientation
of particle $j$ with respect to the reference particle
$i$. Determining this minimal VB for all particles $i$ in the packing
yields a list of $c_{\rm m}$ values for a given $\mathbf{\hat{c}}$ (which is always
relative to the orientation $\mathbf{\hat{t}}_i$). The CDF $P(\mathbf{c},z)$
simply follows by counting the number of values larger than a
specified $c$.

\begin{figure*}[ht]
\begin{center}
\includegraphics[width=16cm]{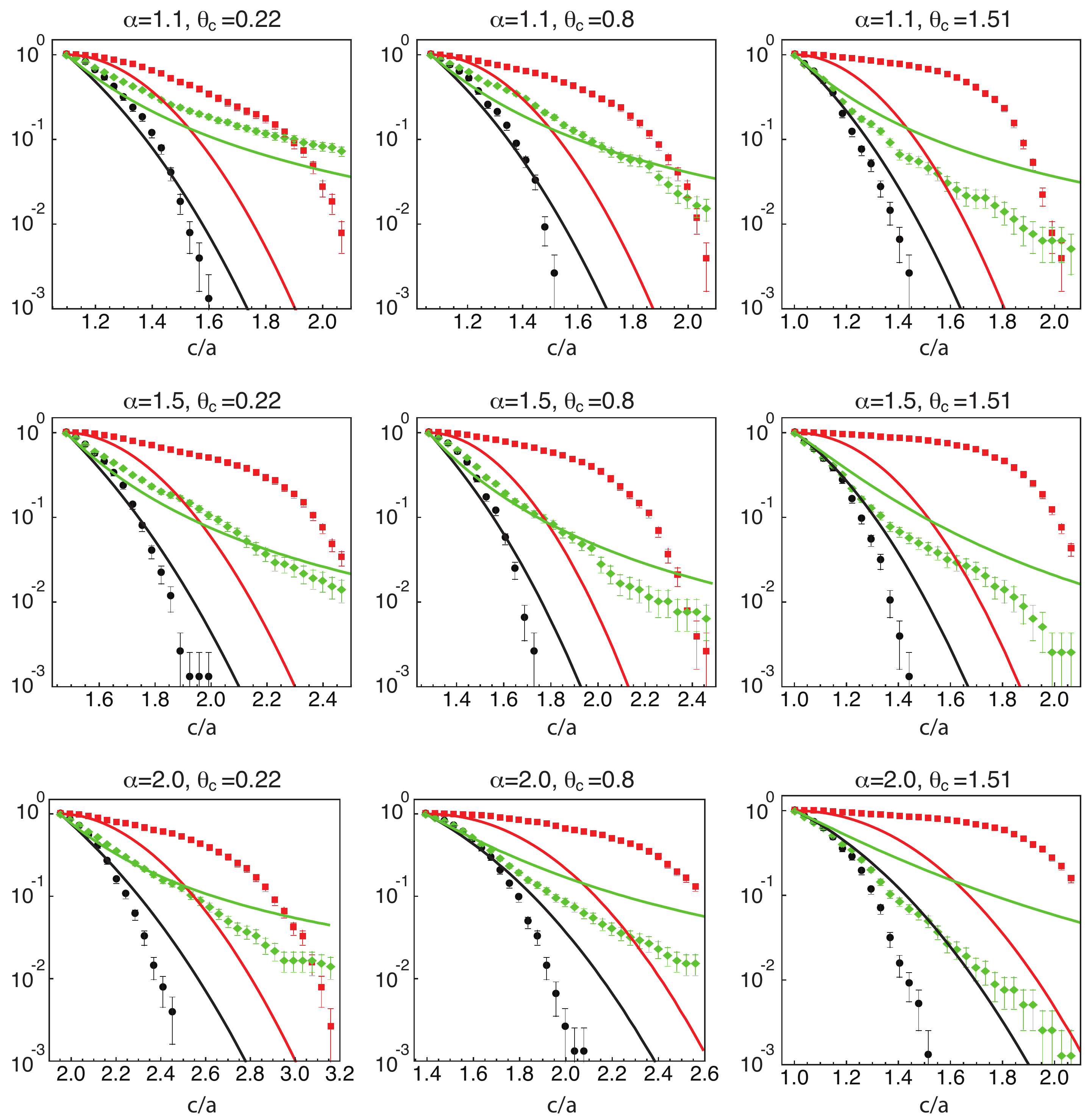} 
\caption{\label{Fig_approx}}
\end{center}
\end{figure*}

Due to the rotational symmetry of the spherocylinders, the
orientational dependence of $P(\mathbf{c},z)$ is reduced to
$P(c,\theta_{\rm c};z)$, where $\theta_{\rm c}$ is the polar angle of the
orientation $\mathbf{\hat{c}}$ in spherical coordinates. Moreover, due to
inversion symmetry it is sufficient to select only
$\theta_{\rm c}\in[0,\pi/2]$. Therefore, we choose three $\theta_{\rm c}$ values
to cover this range: $\theta_{\rm c}=0.22,0.8,1.51$. We also use the
rotational symmetry to improve the sampling of $P(\mathbf{c},z)$: We fix
$\theta_{\rm c}$ to one of the three values, but select a number of
azimuthal angles at random. Since the packing is statistically
isotropic for all azimuthal angles, the resulting $c_{\rm m}$ value for
these directions can all be included in the same ensemble. We consider
three different aspect ratios $\alpha=1.1,1.5,2.0$ of the
spherocylinders to capture a range of different shapes. The results
are plotted in Fig.~\ref{Fig_approx}.

We test the two main approximations considered in the theory: (a) The
derivation of $P(\mathbf{c},z)$ using a liquid like theory of
correlations as done in Refs.~\cite{Song08,Jin10b} leading to the
exponential form of Eq.~(\ref{cdf}). (b) The factorization of this CDF
into contact and bulk contributions as in
Eq. (\ref{fact}). This approximation neglects the correlations between
the contacting particles and the bulk. In Fig.~\ref{Fig_approx}, we
test these approximations by comparing theory and simulations for
three different CDFs: $P(\mathbf{c},z)$, $P_{\rm B}(\mathbf{c})$ and $P_{\rm C}(\mathbf{c},z)$, Eqs.~(\ref{cdf})--(\ref{pb}). In order to determine the $P_{\rm B}(\mathbf{c})$ from the
simulation data we need to take the contact radius
$r^*(\mathbf{\hat{r}}_j,\mathbf{\hat{t}}_j)$ between particle $i$ and
any particle $j$ into account. The minimal VB, $c_{\rm m}$, is determined
from the contributed VBs of particles in the bulk only, i.e.,
particles with
$r_j>r^*(\mathbf{\hat{r}}_j,\mathbf{\hat{t}}_j)$. Likewise, $P_{\rm C}(\mathbf{c})$ is determined from the simulation data by only considering VBs of contacting particles with
$r_j=r^*(\mathbf{\hat{r}}_j,\mathbf{\hat{t}}_j)$.

Following this procedure, we have tested these approximations with the
computer generated packings. We find (Fig.~\ref{Fig_approx}): {\it (i)} The contact term $P_{\rm C}$ is
well approximated by the theory for the full range of $c$; {\it (ii)}
For small values of $c$ the bulk distribution $P_{\rm B}$ is well
approximated by the theory, and deviations are observed for larger
$c$; {\it (iii)} The full CDF $P(\mathbf{c})$ agrees well between the
computer simulations and the theory, especially for small $c$. The
small values of $c$ provide the dominant contribution in the
self-consistent equation to calculate the average Voronoi volume
Eq.~(\ref{integral}), and therefore to the main quantity of interest,
the volume fraction of the packing. This can be seen by rewriting
Eq.~(\ref{integral}) as \be
\label{intconv}
\overline{W}(z)=V_\alpha+\oint\D \mathbf{\hat{c}} \int_{c^*(\mathbf{\hat{c}})}^\infty\D
c \,P(c,\mathbf{\hat{c}};z), \ee since the CDF is trivially unity for $c$
values smaller than the hard-core boundary $c^*(\mathbf{\hat{c}})$. The main
contribution to the integral then comes from $c$ values close to
$c^*(\mathbf{\hat{c}})$ due to the decay of the CDF.

Systematic deviations in our approximations arise in the bulk
distribution $P_{\rm B}$ for larger values of $c$, but, interestingly, the
slope of the decay still agrees with our theory. Overall, the
comparison highlights the mean-field character of our theory: Correlations are captured well up to about the first coordination
  shell of particles, after which theory and simulations diverge,
  especially for the bulk term. The agreement is acceptable for the
  nearest neighbour-shell, but is incorrect for the second neighbours.
Beyond this shell, bulk particles are affected in a finite range by
correlations that we do not address, since we assume a uniform
distribution of the density of these particles; this is a typical
assumption in a mean-field theory. The additional unaccounted
correlations lead to a slightly higher probability to observe the VB
at intermediate $c$ values in the simulation, compared with our
theory. However, these deviations from simulations are small. For
instance, Fig.~\ref{Fig_approx} indicates that for a typical value
$\alpha=1.5$ and polar angle $\theta_S=0.22$, the numerically measured
CDF $P(\mathbf{c},z)$ at a relative large value $c/a=2$ is of the
order of $10^{-3}$, while the theory predicts this probability at a
slightly larger value of $c/a=2.07$. This small discrepancy is not
relevant, since such a value of the probability is negligibly small in
the calculation of the volume fraction in Eq.~(\ref{integral}). Thus,
because of this small probability to find the VB with values larger
than $c/a=2$, the deviations expected from our approximations are
small. These results indicate that, overall, the theory captures the
distribution of VBs in the region of small $c$, which is the
relevant region in the calculation of the volume fraction.

\begin{figure*}[ht]
\begin{center}
\includegraphics[width=18cm]{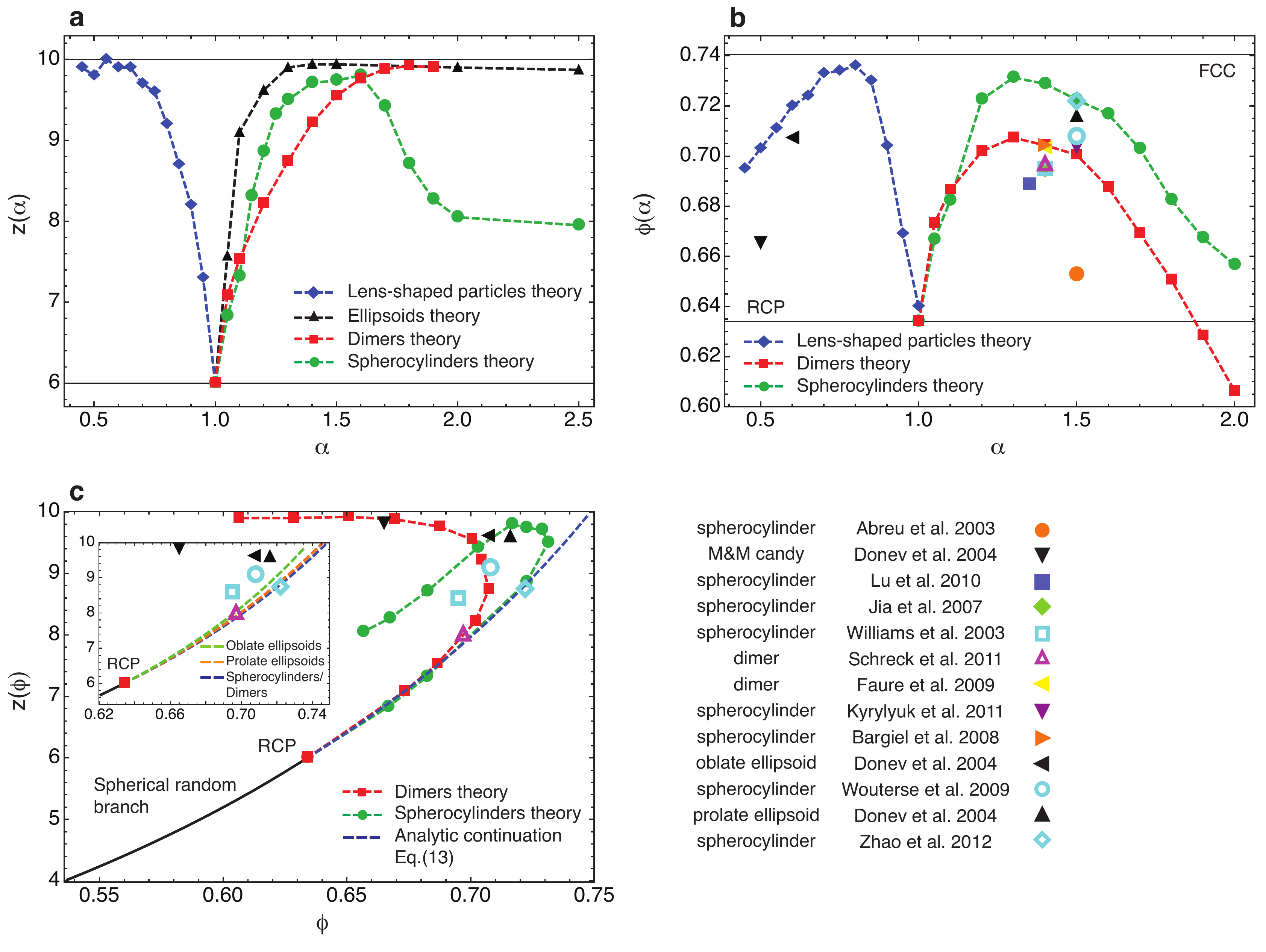}
\caption{\label{Fig_results}}
\end{center}
\end{figure*}

The neglected higher-order correlations in the upper coordination
shells can only decrease the volume fraction in the calculation
leading to smaller packing densities. Following this analysis, we
  interpret our predicted packing fractions as upper bounds for the
  empirically found ones, which is indeed observed in
  Fig.~\ref{Fig_results}b,c.

\vspace{0.5cm}

{\bf Acknowledgements:} We gratefully acknowledge funding by NSF-CMMT
and DOE Office of Basic Energy Sciences, Chemical
Sciences, Geosciences, and Biosciences Division. We are grateful to C.~F. Schreck and C.~S. O'Hern
for discussions and for providing simulated data on 3d packings of
dimers. We are also grateful to F. Potiguar for discussions, T. Zhu
for simulations and M. Danisch for theory. We also thank F. Zamponi,
P. Charbonneau and Y. Jin for discussions on the interpretation of
protocol-dependent packings.

\vspace{0.5cm}

{\bf Author contributions:} AB, RM, LB, LP, and HAM designed research, performed research, and wrote the paper.

\vspace{0.5cm}

{\bf Competing financial interests:} The authors declare no competing
financial interests.

\clearpage

\begin{figure*}[ht]
\begin{center}
\includegraphics[width=6cm]{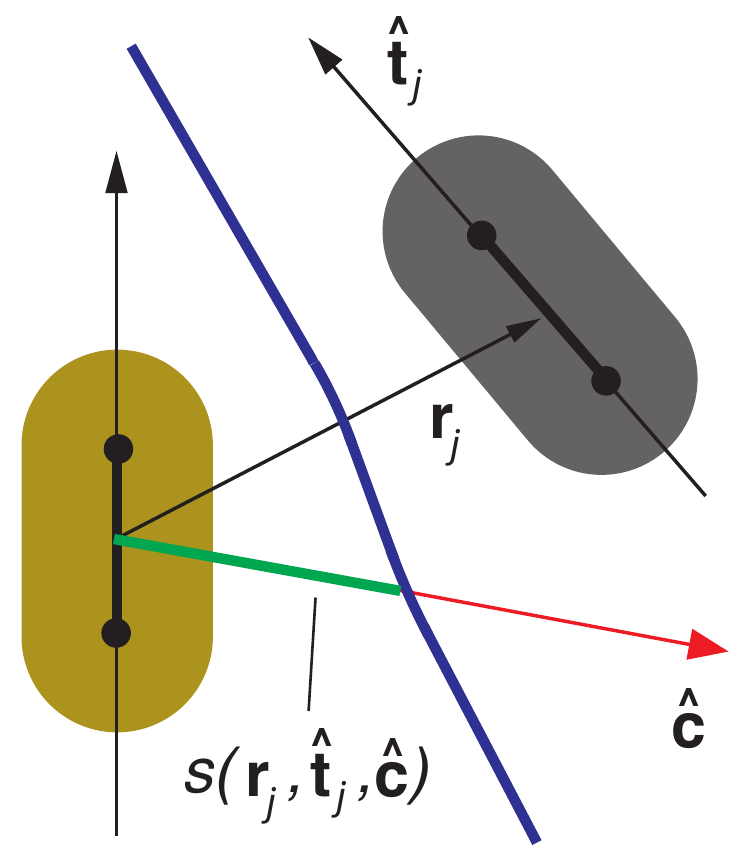}
\caption{\label{Fig_coord}}
\end{center}
\end{figure*}

FIG.~\ref{Fig_coord}: {\bf Parametrization of the Voronoi
  boundary}. The Voronoi boundary (VB) in blue, denoted by $s(\mathbf{r}_j,\mathbf{\hat{t}}_j,\mathbf{\hat{c}})$, along a direction $\mathbf{\hat{c}}$ between two spherocylinders of relative
position $\mathbf{r}_j$ and orientation $\mathbf{\hat{t}}_j$.

\begin{figure*}[ht]
\begin{center}
\includegraphics[width=12cm]{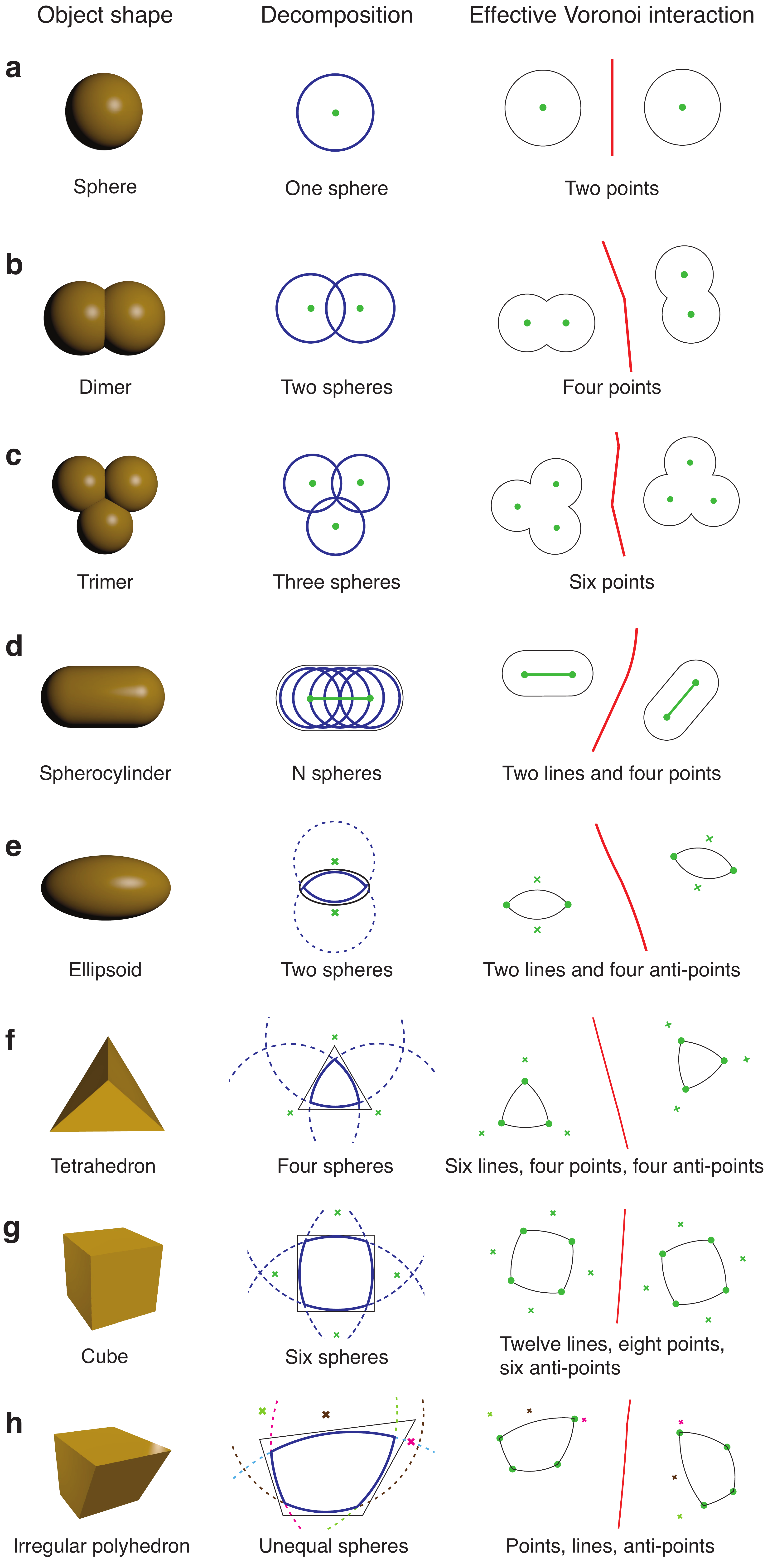}
\caption{\label{Fig_shapes}}
\end{center}
\end{figure*}

FIG.~\ref{Fig_shapes}: {\bf Decomposition of various shapes and
  effective Voronoi interactions}. Arbitrary object shapes can be
decomposed into unions and intersections of spheres. {\bf (a)--(d)} Union of spheres. The VB between two such objects is equivalent to the VB between the point multiplets at the centre
of the spheres, as shown for four basic shapes. {\bf (e)--(g)} Intersection of spheres.
The VB between such intersections is
equivalent to that between multiplets of ``anti-points" at the center
of the spheres, indicated by crosses, and, in addition, lines at the
edges of the intersections, shown as points in (e)--(d). The additional lines arise due to the
positive curvature at the singular intersections, resulting in edges
that point outwards from the particle rather than inwards. In the case
of dimers and trimers shown in (b) and (c), the curvature
is negative and the edges do not influence the VB. The
generalization to (f) tetrahedra-like, (g) cubes, and (h) irregular polyhedra-like shapes is straightforward. Note that
the VBs drawn in (e)--(h) are only qualitative.

\begin{figure*}[ht]
\begin{center}
\includegraphics[width=12cm]{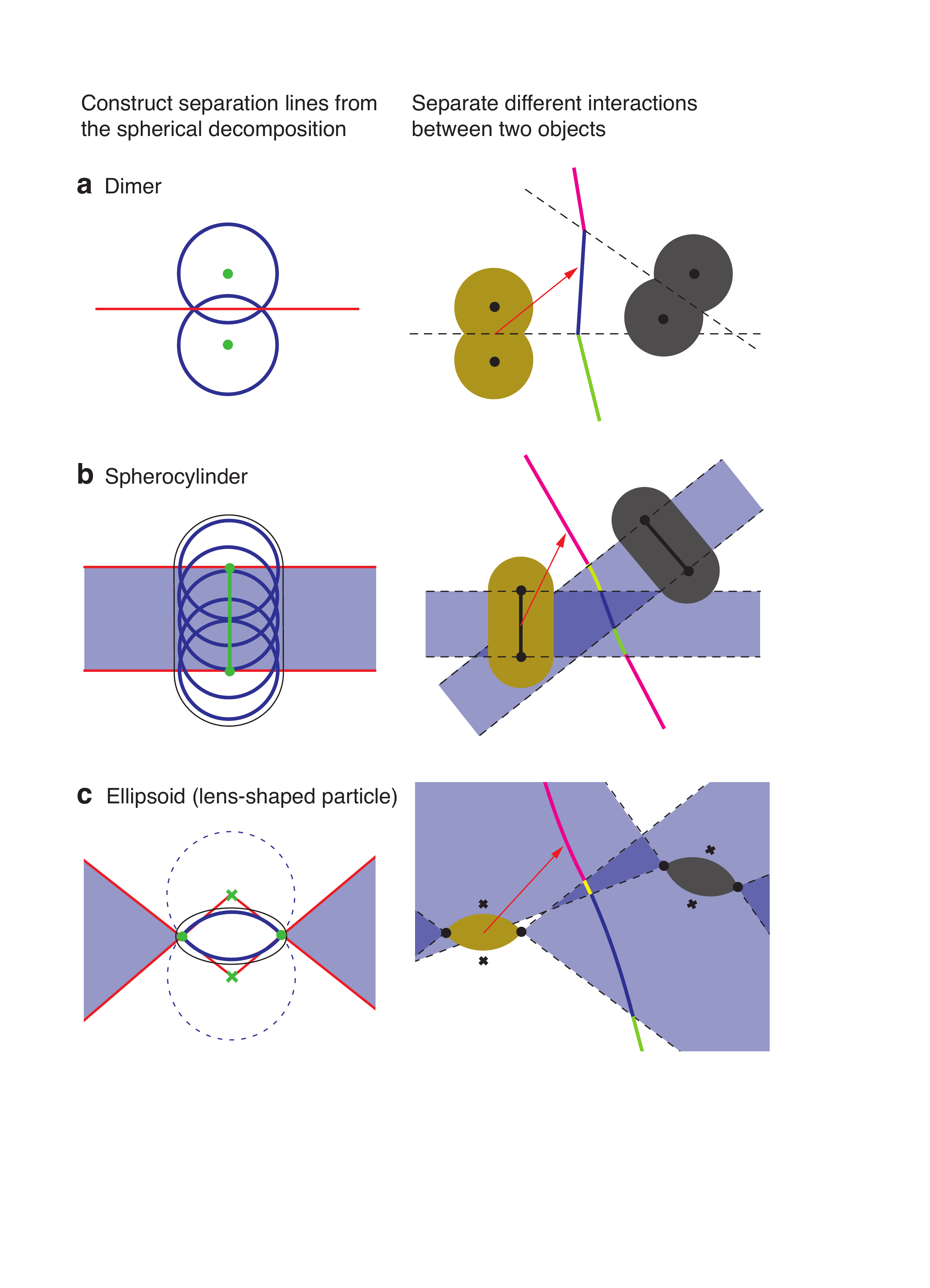}
\caption{\label{Fig_algorithms}}
\end{center}
\end{figure*}

FIG.~\ref{Fig_algorithms}: {\bf Analytical solution to determine the
  VB for non-spherical objects}. {\bf (a)} The VB between two objects of a given
relative position and orientation consists of the VBs between
particular spheres on each of the two objects. The spheres that
interact are determined by separation lines given as the VBs between the spheres in
the filling. For dimers, there is
one separation line for each object, tesselating space into four
areas, in which only one interaction is correct. The pink part in (a), e.g., is the VB between the two upper spheres. 
{\bf (b)} The dense overlap of
spheres in spherocylinders leads to a line as effective Voronoi
interaction at the centre of the cylindrical part. This line
interaction has to be separated from the point interactions due to the
centres of the spherical caps as indicated. Overall, the two
separation lines for each object lead to a tessellation of space into
nine different areas, where only one of the possible line-line,
line-point, point-line, and point-point interactions is possible. The yellow part in (b), e.g.,
is due to the upper point on spherocylinder 1 and the line of 2. Regions of line interactions are indicated by blue shades. {\bf
  (c)} The spherical decomposition of ellipsoid-like shapes is analogous to dimers, only that now the opposite sphere centres
interact. We indicate this inverted interaction by a cross at the
centres of the spheres and refer to these points as ``anti-points". In
addition, the positive curvature at the intersection point leads to an
additional line interaction, which is a circle in 3d (a point in 2d)
and indicated here by two points. The separation lines are then given
by radial vectors through the intersection point/line. The Voronoi
interaction between two ellipsoids is thus given by two pairs of two
anti-points and a line, which is the same class of interactions as
spherocylinders. The different point and line interactions are
separated analogous to spherocylinders, as shown.

\begin{figure*}[ht]
\begin{center}
\includegraphics[width=12cm]{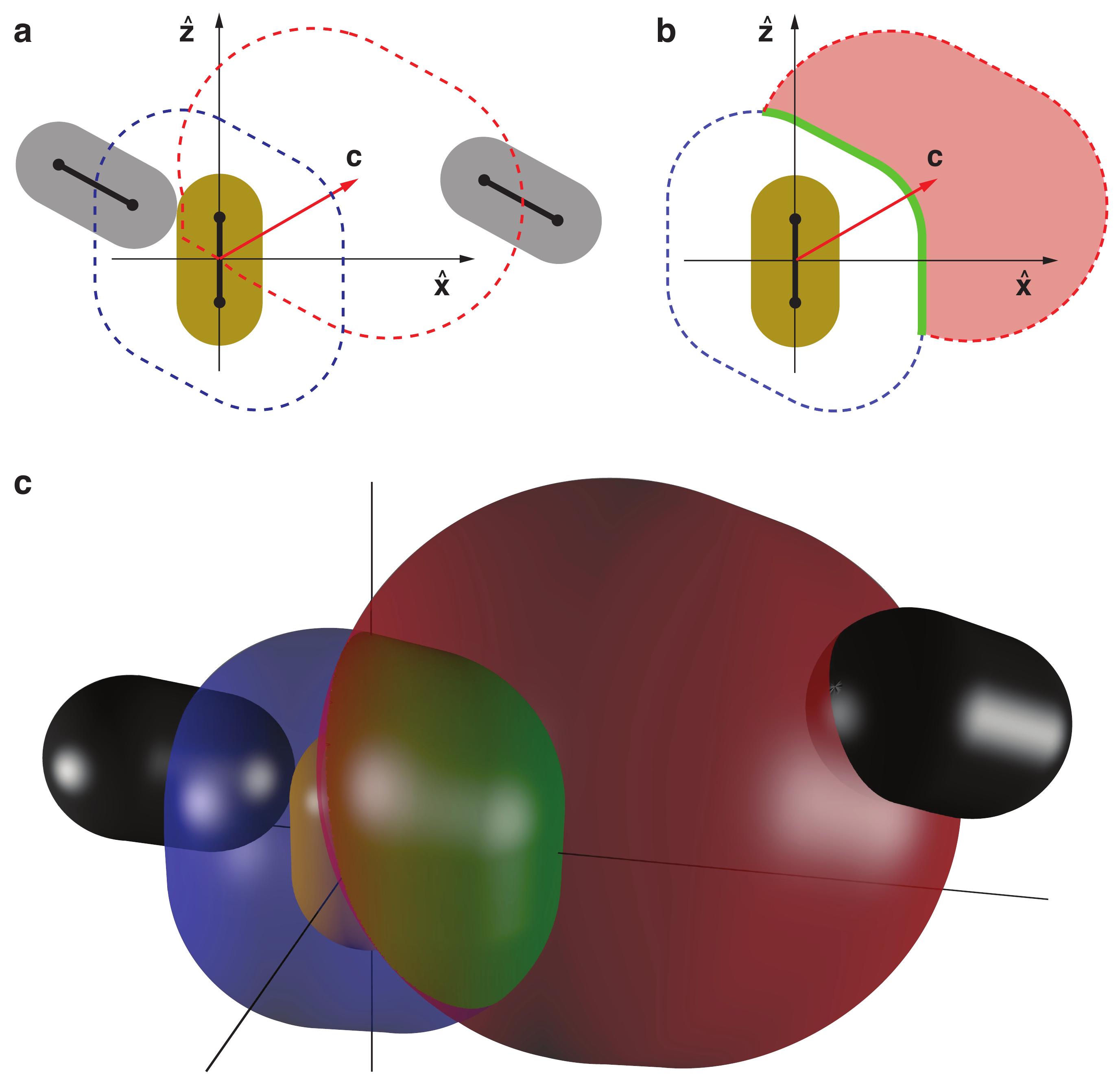}
\caption{\label{Fig_volumes}}
\end{center}
\end{figure*}

FIG.~\ref{Fig_volumes}: {\bf The Voronoi excluded volume and
  surface}. {\bf (a)} The hard-core repulsion between two objects
defines the hard-core excluded volume $V_{\rm ex}$ (enclosed by a
dashed blue line): This volume is excluded for the centre of mass of
any other object. Packings of rods in the limit $\alpha\to\infty$ can be described
by a simple random contact equation based on $V_{\rm ex}$
\cite{Philipse96}. We introduce the Voronoi excluded volume $\Omega$
(enclosed by a dashed red line), which is the basis of our statistical
theory of the Voronoi volume. The volume $\Omega$,
Eq.~(\ref{methods_vex}), is excluded by the condition that no other
particle should contribute a VB smaller than $c$ in the direction
$\mathbf{\hat{c}}$, which defines the CDF $P(\mathbf{c},z)$. {\bf (b)}
Taking into account the hard-core exclusion leads to the effective
Voronoi excluded volume $V^*$ (indicated as red volume), which is
excluded for bulk particles. Likewise, the overlap of $V_{\rm ex}$ and
$\Omega$ excludes the surface $S^*$ (thick green line) for all
contacting particles. The volumes are shown here for a single
orientation $\mathbf{\hat{t}}$. {\bf (c)} The 3d plot corresponding to
(b): The central particle is in brown, $V_{\rm ex}$ is indicated
in blue, $V^*$ in red, and $S^*$ in green.

\begin{figure*}[ht]
\begin{center}
\includegraphics[width=12cm]{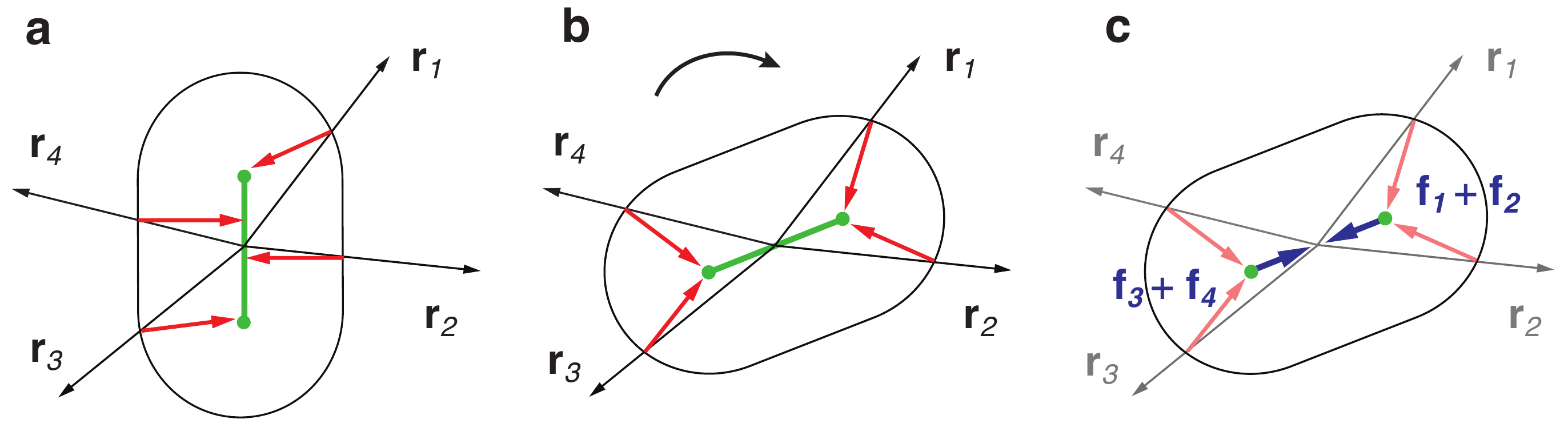}
\caption{\label{Fig_spheroz}}
\end{center}
\end{figure*}

FIG.~\ref{Fig_spheroz}: {\bf Quantitative method to calculate
  $z(\alpha)$}. {\bf (a)} A two-dimensional sketch of a spherocylinder
with a random configuration of contact directions $\mathbf{r}_j$. The
associated forces are along directions $\hat{n}_j$ normal to the
surface (indicated in red) and torques are along $\mathbf{r}_j\times
\mathbf{\hat{n}}_j$. From these directions, one can determine if
mechanical equilibrium has some redundancy, i.e., if force and torque
balance equations are not linearly independent. The configuration
shown has no redundancy: The equivalent situation in three dimensions
would show force and torque balance equations as five different
constraints (the most general case for a three-dimensional particle
would be six constraints, but the torque along the axis of a
spherocylinder is always vanishing, due to its rotational
symmetry). {\bf (b)} Here, the spherocylinder is rotated. With the
same contact directions $\mathbf{r}_j$ as in (a), the contact
force directions $\mathbf{\hat{n}}_j$ are now modified. We explore the
space of possible orientations for the spherocylinder, and try to find
configurations which maximize redundancy in the mechanical equilibrium
conditions. {\bf (c)} As an example, this orientation exhibits some
redundancy: All the contacts are on the spherical caps of the
spherocylinder. Therefore, $\mathbf{f}_1+\mathbf{f}_2$ and
$\mathbf{f}_3+\mathbf{f}_4$ are aligned with the spherocylinder axis
and the condition of force balance automatically implies torque
balance. If this is the orientation of the spherocylinder for which
redundancy is maximal, we associate the number of linearly independent
equations (i.e., the effective number of degrees of freedom) from the
mechanical equilibrium condition with the set of contact directions
$\{\mathbf{r}_j \}$ and perform an average over the possible sets of
$\{\mathbf{r}_j \}$. This yields the averaged effective number of
degrees of freedom $\tilde{d}_{\rm f}(\alpha)$ for a spherocylinder having
an aspect ratio $\alpha$ and the coordination number follows as
$z(\alpha)=2\tilde{d}_{\rm f}(\alpha)$. Note that for non-convex shapes like
dimers, the resulting $z$ is the number of contacting neighbours, not
the number of contacts, which can exceed the former.

FIG.~\ref{Fig_results}: {\bf Theoretical predictions for packings of
  dimers, spherocylinders and lens-shaped particles}. {\bf (a)} The
function $z(\alpha)$ determined by evaluating the probability of
degenerate configurations. Both
spherocylinders and dimers increase up to just below the isostatic
value $z=10$. For dimers, $z(\alpha)$ is the number of contacting
neighbours, not the number of contacts, since a single contacting
particle can have more than one contacting point. For
spherocylinders, $z$ reduces to $8$ for large $\alpha$, since the
forces acting on the cylindrical part are coplanar and reduce the
effective degree of freedom. We also include the results from our
method for prolate ellipsoids of revolution and lens-shaped
particles. {\bf (b)} The predicted packing
fraction $\phi(\alpha)$ of spherocylinders, dimers, and lens-shaped
particles compared with simulation results of maximal densities from
the literature. We predict the maximal packing fraction of
spherocylinders $\phi_{\rm max}=0.731$ at $\alpha=1.3$ and of dimers
$\phi_{\rm max}=0.707$ at $\alpha= 1.3$, demonstrating that
spherocylinders pack better than dimers. For the lens-shaped particles
we obtain $\phi_{\rm max}=0.736$ at $\alpha= 0.8$. {\bf (c)} By
plotting $z$ vs $\phi$ we obtain a phase diagram for smooth shapes. We
observe that the spherical random branch $\phi_{\rm sph}$, which ends
at the RCP point at $(0.634,6)$ \cite{Song08}, in fact continues
smoothly upon deformation into dimers and spherocylinders as predicted
by our theory. The spherocylinder continuation provides
a boundary for all known packing states of rotationally symmetric
shapes. Inset: The continuations from RCP. For a given value of $z$, the densest
packing is achieved by spherocylinders, followed by dimers, prolate
ellipsoids, and oblate ellipsoids. Note that the continuations for
spherocylinders and dimers are almost identical.

FIG.~\ref{Fig_approx}: {\bf Comparison of the CDF with simulation
  data.} We plot the theoretical predictions (solid lines) for $P(\mathbf{c},z)$ (black), $P_{\rm B}(\mathbf{c})$ (red), and $P_{\rm C}(\mathbf{c},z)$ (green) with the corresponding CDFs sampled
from simulated configurations (symbols) of spherocylinders. For each
aspect ratio $\alpha=1.1,1.5,2.0$ we plot results for three values of
the polar angle $\theta_{\rm c}\in[0,\pi/2]$. We generally observe that the
three CDFs agree quite well in the regime of small $c$ values, which
provides the dominant contribution to the average Voronoi volume $\overline{W}(z)$. The same plots are shown on a linear scale in the Supplementary Figure S1. The error bars denote the root mean square error of the finite-size sampling.

\newpage

\begin{table*}[ht]
\centering
\begin{tabular}{c | c | c | c }
Shape & $\qquad \phi_{\rm max} \qquad$ & Aspect ratio at $\phi_{\rm max}$ & Reported $z$ \\ 
\hline 
spherocylinder$^5$ & 0.653 & 1.5 &  \\
M\&M candy$^6$ & 0.665 & 0.5 & 9.8  \\
spherocylinder$^{14}$  & 0.689 & 1.35 &   \\
spherocylinder$^8$ & 0.694 & 1.4 &  \\
spherocylinder$^4$ & 0.695 & 1.4 & 8.6  \\
dimer & 0.697 & 1.4 & 8.0   \\
dimer$^{12}$ & 0.703 & 1.4 &  \\
spherocylinder$^{15}$ & 0.703 & 1.5 &  \\
spherocylinder$^9$ & 0.704 & 1.4 & \\
oblate ellipsoid$^6$ & 0.707 & 0.6 & 9.6  \\
{\bf dimer(theory)} & 0.707 & 1.3 & 8.74  \\
spherocylinder$^{10}$ & 0.708 & 1.5 & 9.1  \\
prolate ellipsoid$^6$ & 0.716 & 1.5 & 9.6  \\
spherocylinder$^{17}$ & 0.722 & 1.5 & 8.7  \\
{\bf spherocylinder (theory)} & 0.731 & 1.3 & 9.5  \\
{\bf lens-shaped particle (theory)} & 0.736 & 0.8 & 9.2  \\
\hline
general ellipsoid$^6$ & 0.735 & &  \\
general ellipsoid$^7$ & 0.74 & & 10.7 \\
tetrahedron$^{13}$ & 0.76 & & 12  \\
tetrahedron$^{16}$ & 0.763 & &  \\
tetrahedron$^{11}$ & 0.7858 & & 
\end{tabular}
\caption{\label{table1}{\bf Overview of packing fractions from simulations and experiments}. The maximal packing fraction $\phi_{\rm max}$
  and reported coordination number $z$ at $\phi_{\rm max}$ of random
  packings of spherocylinders, dimers, ellipsoids and tetrahedra,
  determined from simulations and experiments. The aspect ratio is
  defined for rotationally symmetric objects. Some simulations do not
  report $z$. Results are separated by the symmetry of the object
  (rotationally symmetric and asymmetric) and ordered by packing
  fraction. From the available empirical data we cannot conclude
  whether spherocylinders pack better than dimers or ellipsoids of
  revolution, for instance.}
\end{table*}

\newpage

\newpage

\newpage

\newpage

\newpage

\clearpage

\newpage

\newpage

\section*{\large Supplementary Information: Mean-field theory of random close packings of axisymmetric particles}

\begin{figure*}[ht]
\begin{center}
\includegraphics[width=16cm]{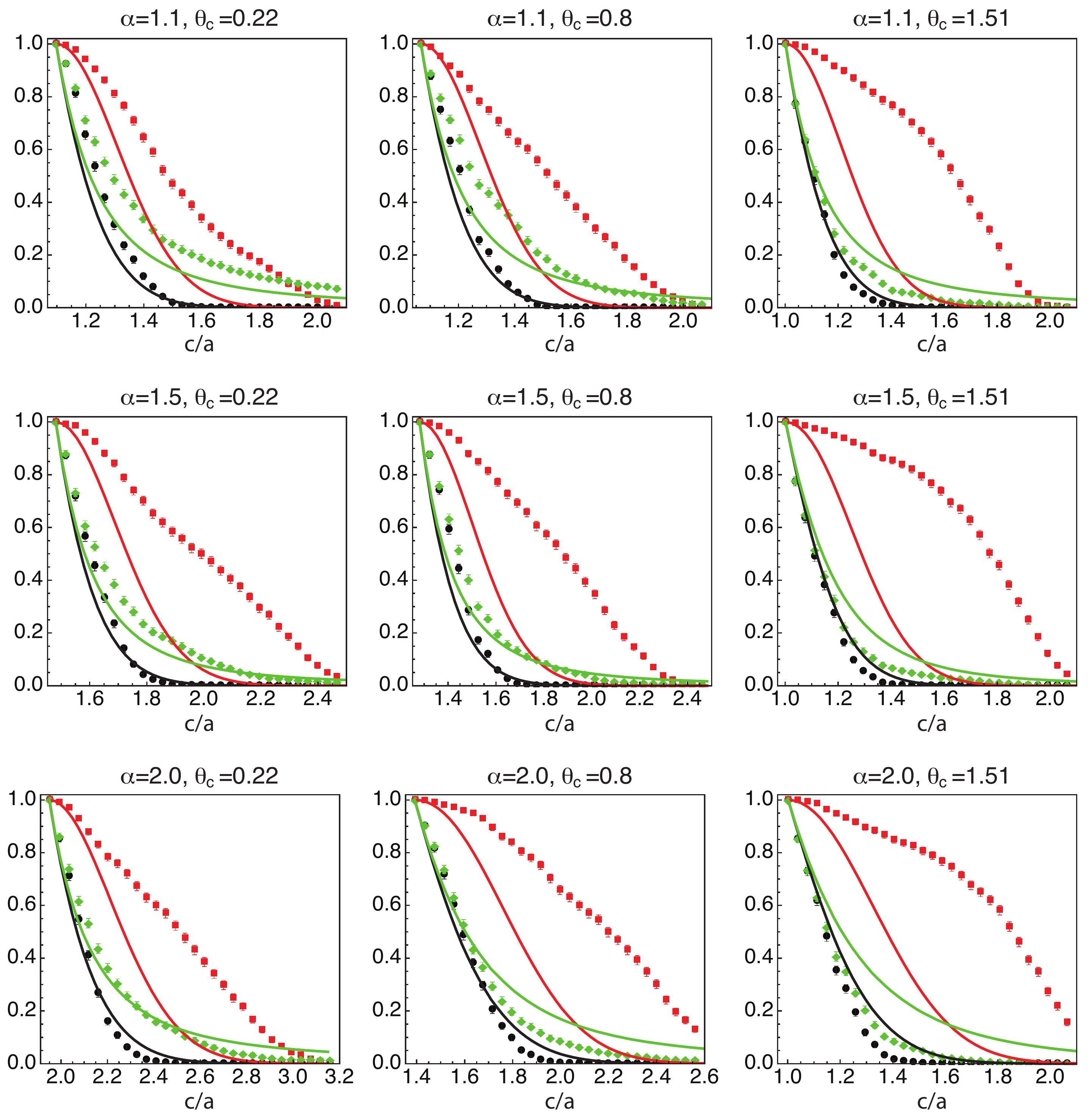} 
\caption{{\bf The plots of Fig.~\ref{Fig_approx} shown on a linear scale.} We plot the theoretical predictions (solid lines) for $P(\mathbf{c},z)$ (black), $P_{\rm B}(\mathbf{c})$ (red), and $P_{\rm C}(\mathbf{c},z)$ (green) with the corresponding CDFs sampled from simulated configurations (symbols) of spherocylinders. For each aspect ratio $\alpha=1.1,1.5,2.0$ we plot results for three values of the polar angle $\theta_{\rm c}\in[0,\pi/2]$. We generally observe that the three CDFs agree quite well in the regime of small $c$ values, which provides the dominant contribution to the average Voronoi volume $\overline{W}(z)$. The error bars denote the root mean square error of the finite-size sampling.}
\end{center}
\end{figure*}

\clearpage

\begin{figure*}[ht]
\begin{center}
\includegraphics[width=10cm]{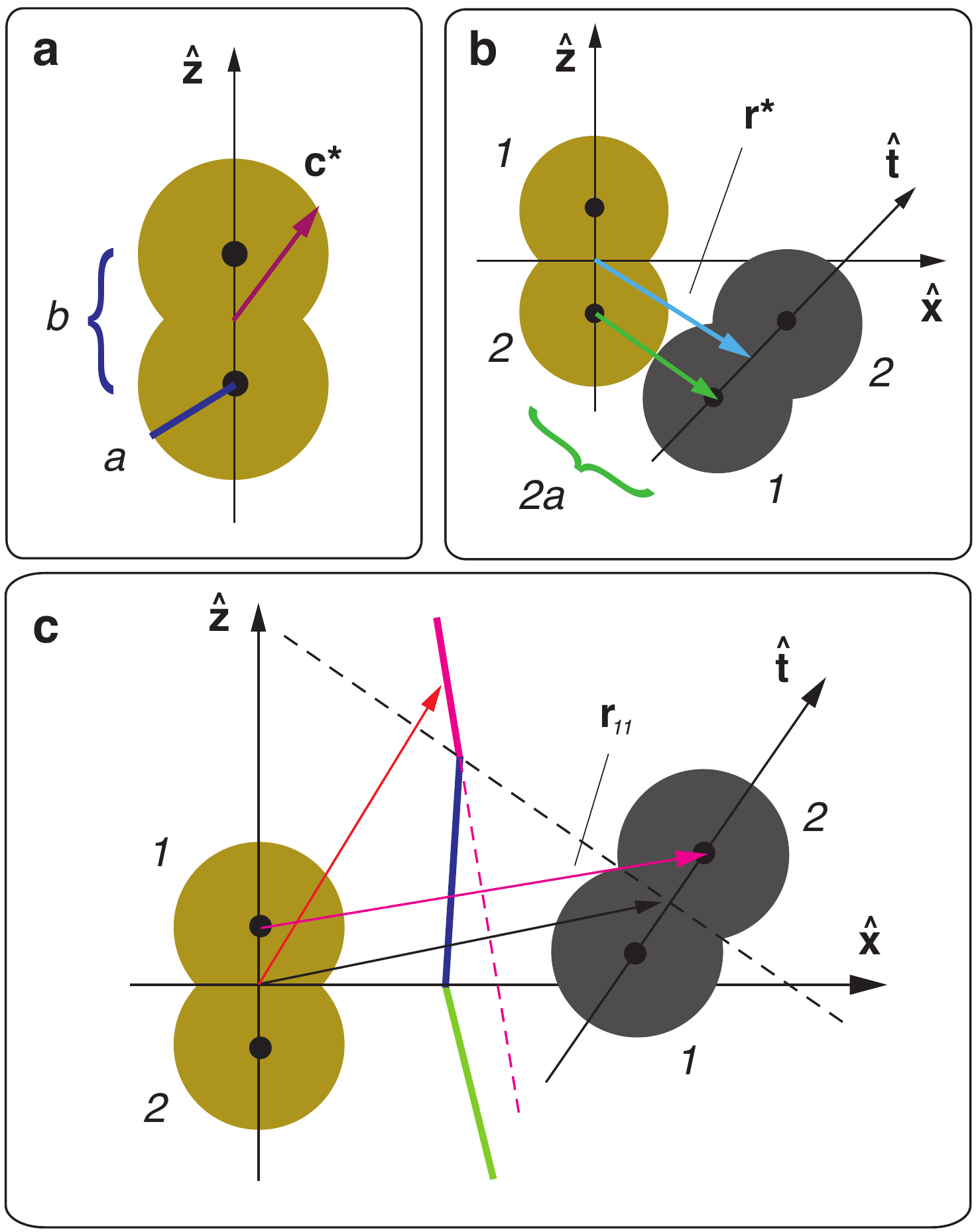}
\caption{\label{SFig_dimers} {\bf Parametrization of dimers.} {\bf (a)} A dimer with parameters $a$ and $b$. The hard core boundary is parametrized by the vector $\mathbf{c}^*=c^*(\theta_{\rm c})\mathbf{\hat{c}}$. {\bf (b)}, The contact radius $r^*(\mathbf{\hat{r}},\mathbf{\hat{t}})$ (light blue) is determined by the condition of contact between sphere $2$ on the $i$th particle and $2$ on the $j$th: $r^*=r^*_{22}$, where $r^*_{22}$ is given by Eq.~(\ref{r22eq}). {\bf (c)} The VB between two dimers of relative orientation $\mathbf{\hat{t}}$ and position $\mathbf{r}$. The VB is determined by the interactions between the different point pairs (indicated in different colors), which are separated following our algorithm in Fig.~\ref{Fig_algorithms}a. The pink part of the VB, e.g., is the VB between points $1$ and $1$, and is given by Eq.~(\ref{s11vor}).}
\end{center}
\end{figure*}

\clearpage

\begin{figure*}[ht]
\begin{center}
\includegraphics[width=14cm]{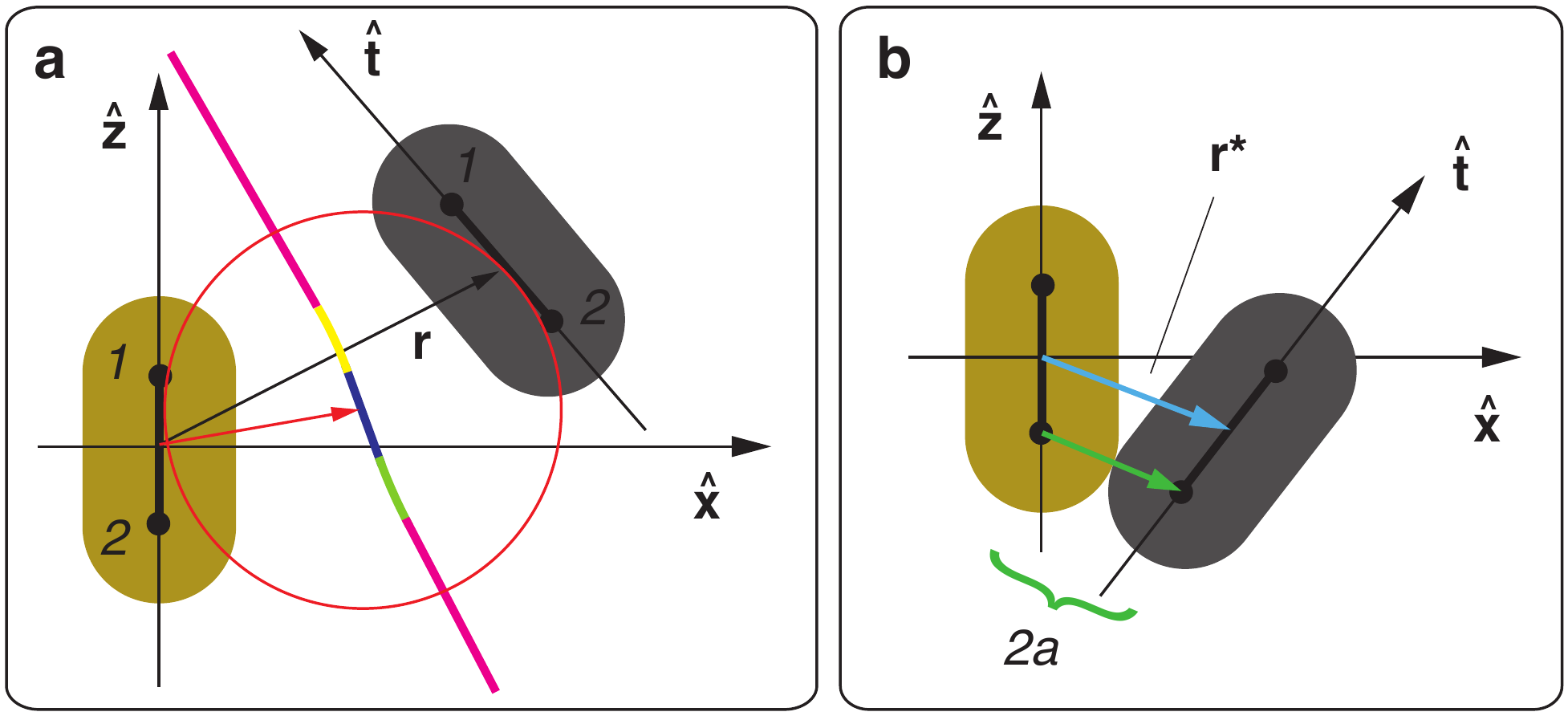} 
\caption{\label{SFig_spheros}{\bf Parametrization of spherocylinders.} {\bf (a)} The VB between two spherocylinders of relative orientation $\mathbf{\hat{t}}$ and position $\mathbf{r}$. The VB consists of the VBs due to the interaction of the four points and two lines (indicated in different colors), which are separated following our algorithm in Fig.~\ref{Fig_algorithms}a. The blue part of the VB, e.g., is due to the line-line interaction given by Eq.~(\ref{lleq}): a sphere centered on this part touches the rods $i$ and $j$ for a unique radius. {\bf (b)} The contact radius $r^*(\mathbf{\hat{r}},\mathbf{\hat{t}})$ for two spherocylinders. Here, the contact is due to the spherical endcaps.
}
\end{center}
\end{figure*}

\clearpage

\begin{table}[ht]
\centering
\begin{tabular}{c | c | c | c}
Shape & $\qquad M_z \qquad$ & $\qquad M_{\rm b} \qquad$ & $\qquad M_{\rm v} \qquad$ \\ 
\hline 
$\quad$spherocylinder$\quad$ & 2.767 & 1/2 & 3/2  \\ 
dimer & 3.60 & 1/2 & 3/2  \\ 
prolate ellipsoid & 4.833 & 1/3 & 1  \\ 
oblate ellipsoid & -5.167 & 1/3 & 1  
\end{tabular}
\caption{\label{suptable}{\bf Values of the shape-dependent constants in the analytic continuation of RCP}. Note that the $M_z$ values for dimers and
  spherocylinders are taken from Fig.~\ref{Fig_results}a in the main text, and the ones
  for the two rotationally symmetric ellipsoids from
  Ref.~\cite{Donev04}. Due to the limited data, $M_z$ is determined by
  linear interpolation.}
\end{table}

\clearpage

\section*{Supplementary Note 1}

In our simulation, we treat the case of spherocylinders. The
interaction force between two particles is described as a linear
function of the overlap. To obtain a jammed configuration, we follow
previously studied protocols \cite{Song08}. We start our simulation
with a number of particles $N$, and generate particle positions
randomly within a cubic box with size $L$ and periodic boundary
conditions. We first compress the initial system by shrinking the box
size $L$ to reach a certain pressure (which is very high at the first
step, $P=10^6$) and then let it relax fast until it fails to jam. We
then compress the system and relax repeatedly several times until the
system ends up in a stable but overcompressed configuration. This
means that the pressure limit and relaxation rate we choose are too
high to get to the jamming point. Thus we lower the pressure limit and
relax the system slowly to obtain a less overcompressed
configuration. We tune the two parameters, pressure and relaxation
rate, until the system reaches a well jammed configuration with a very
low pressure ($P<10$). This procedure brings the system to the jamming
point with minimal overlap.

\clearpage

\section*{Supplementary Methods}

\section*{Calculation of the Voronoi boundary and the contact radius for dimers and spherocylinders}

\label{Secsupp_calc}

The Voronoi boundary (VB) between two objects is defined as the
hypersurface that contains all the points that are equidistant to both
objects. As before, we set the centre of our coordinate system to the
centre of mass of particle $i$ and fix the orientation of this
particle along $\mathbf{\mathbf{\hat{z}}}$. Given a direction $\mathbf{\hat{c}}$, a point on the
VB is found at $s\mathbf{\hat{c}}$, where $s$ depends on the position $\mathbf{r}$ and
orientation $\mathbf{\hat{t}}$ of particle $j$: $s=s(\mathbf{r},\mathbf{\hat{t}},\mathbf{\hat{c}})$. The value
of $s$ is obtained from two conditions:
\begin{enumerate}
\item The point $s\mathbf{\hat{c}}$ has the minimal distance to each of the two objects along the direction $\mathbf{\hat{c}}$.
\item Both distances are the same.
\end{enumerate}
The VB between two spheres of equal radii is the same as the VB between two points at the centres of the spheres. Therefore, condition $1$ is trivially satisfied for every $s$ and condition $2$ translates into the equation
\be
\label{VBspherecond}
(s\mathbf{\hat{c}})^2=(s\mathbf{\hat{c}}-\mathbf{r})^2,
\ee
leading to
\be
\label{VBsphere}
s=\frac{r}{2\mathbf{\hat{c}}\mathbf{\hat{r}}},
\ee
i.e., the VB is the plane perpendicular to the separation vector $\mathbf{r}$ at half the separation (see Fig.~\ref{Fig_shapes}a, main text). Already for two spheres of unequal radii, the VB is a curved surface. Taking into account the different radii $a_i$ and $a_j$, Eq.~(\ref{VBspherecond}) becomes
\be
s-a_i=\sqrt{(s\mathbf{\hat{c}}-\mathbf{r})^2}-a_j,
\ee
which has the solution
\be
s=\frac{1}{2}\frac{r^2-(a_i-a_j)^2}{\mathbf{\hat{c}}\mathbf{\hat{r}}-(a_i-a_j)}.
\ee
Finding a solution for both conditions for general non-spherical objects is non-trivial. As discussed in the main part of the paper, from these two building blocks the VB between arbitrarily shaped objects can be constructed following our algorithm in Fig.~\ref{Fig_algorithms} of the main text. For shapes consisting of a dense overlap of equal spheres like spherocylinders, this approach can be simplified by introducing a line interaction: The VB between two spherocylinders is equivalent to the VB between two lines at the centre of the cylindrical part. We first discuss the VB between two dimers, which represents the next simplest shape after a sphere.

\section*{Dimers}

\label{Secsupp_calc_dimer}

A dimer consists of two overlapping spheres and is defined by two parameters: the sphere radius $a$ and the separation of the two sphere centres $b$ (Supplementary Fig.~\ref{SFig_dimers}a). The aspect ratio is then $\alpha=1+b/(2a)$. Due to the rotational symmetry, the hard core boundary $c^*(\mathbf{\hat{c}})$ of a dimer is parameterized by the polar angle $\theta_{\rm c}$ only
\be
\label{cs_dim}
c^*(\mathbf{\hat{c}})=c^*(\theta_{\rm c})=a\left(\tilde{\alpha}|\cos(\theta_{\rm c})|+\sqrt{1-\tilde{\alpha}\sin(\theta_{\rm c})}\right),
\ee
where $\tilde{\alpha}=\alpha-1=b/(2a)$.

The VB between two dimers is generated by four different point interactions, which lead to four different values of the VB for a given direction $\mathbf{\hat{c}}$. In order to determine each of the four VBs in our coordinate system, we need the separation vectors for the four different  point pairs. These are
\be
\label{dimercontacts}
\begin{array}{ccc}
\mathbf{r}_{11}=\mathbf{r}-\frac{b}{2}(\mathbf{\hat{t}}+\mathbf{\hat{z}}),&\qquad& \mathbf{r}_{12}=\mathbf{r}+\frac{b}{2}(\mathbf{\hat{t}}-\mathbf{\hat{z}}),\\
\\
\mathbf{r}_{21}=\mathbf{r}-\frac{b}{2}(\mathbf{\hat{t}}-\mathbf{\hat{z}}),& &\mathbf{r}_{22}=\mathbf{r}+\frac{b}{2}(\mathbf{\hat{t}}+\mathbf{\hat{z}}),
\end{array}
\ee
where the subscript $11$ denotes the top point on the $i$th dimer and the top point on the $j$th dimer. The VB $s\mathbf{\hat{c}}_{11}=s_{11}\mathbf{\hat{c}}$ due to the interaction between points $1$ of $i$ and $1$ of $j$ is then determined from the condition (Supplementary Fig.~\ref{SFig_dimers}c)
\be
\label{s11vor}
\left(s_{11}\mathbf{\hat{c}}-\frac{b}{2}\mathbf{\hat{z}}\right)\mathbf{\hat{r}}_{11}=\frac{r_{11}}{2}.
\ee
Likewise, for $s_{12}\mathbf{\hat{c}}$, $s_{21}\mathbf{\hat{c}}$, and $s_{22}\mathbf{\hat{c}}$. This leads to the four values
\be
\begin{array}{ccc}
s_{11}=\frac{r_{11}}{2\mathbf{\hat{c}}\mathbf{\hat{r}}_{11}}+\frac{b}{2}\frac{\mathbf{\hat{z}}\mathbf{\hat{r}}_{11}}{\mathbf{\hat{c}}\mathbf{\hat{r}}_{11}},&\qquad&s_{12}=\frac{r_{12}}{2\mathbf{\hat{c}}\mathbf{\hat{r}}_{12}}+\frac{b}{2}\frac{\mathbf{\hat{z}}\mathbf{\hat{r}}_{12}}{\mathbf{\hat{c}}\mathbf{\hat{r}}_{12}},\\
\\
s_{21}=\frac{r_{21}}{2\mathbf{\hat{c}}\mathbf{\hat{r}}_{21}}-\frac{b}{2}\frac{\mathbf{\hat{z}}\mathbf{\hat{r}}_{21}}{\mathbf{\hat{c}}\mathbf{\hat{r}}_{21}},& &s_{22}=\frac{r_{22}}{2\mathbf{\hat{c}}\mathbf{\hat{r}}_{22}}-\frac{b}{2}\frac{\mathbf{\hat{z}}\mathbf{\hat{r}}_{22}}{\mathbf{\hat{c}}\mathbf{\hat{r}}_{22}}.
\end{array}
\ee
The VB between the two dimers is then given by $s_{11}$, if the point $s_{11}\mathbf{\hat{c}}$ is inside the appropriate region outlined by the separation lines in Fig.~\ref{Fig_algorithms}a in the main text. This is the case if
\be
\label{dimervor_sep}
s_{11}\mathbf{\hat{c}}\mathbf{\hat{z}}>0,\qquad{\rm and}\qquad \mathbf{\hat{t}}(s_{11}\mathbf{\hat{c}}-\mathbf{r})>0.
\ee
For $s_{12}$ the conditions are
\be
s_{12}\mathbf{\hat{c}}\mathbf{\hat{z}}>0,\qquad{\rm and}\qquad \mathbf{\hat{t}}(s_{12}\mathbf{\hat{c}}-\mathbf{r})<0,
\ee
and likewise for $s_{21}$, and $s_{22}$
\be
s_{21}\mathbf{\hat{c}}\mathbf{\hat{z}}&<&0,\qquad{\rm and}\qquad \mathbf{\hat{t}}(s_{12}\mathbf{\hat{c}}-\mathbf{r})>0,\\
s_{22}\mathbf{\hat{c}}\mathbf{\hat{z}}&<&0,\qquad{\rm and}\qquad \mathbf{\hat{t}}(s_{12}\mathbf{\hat{c}}-\mathbf{r})<0.
\ee
This yields a unique value $s$ for the VB along $\mathbf{\hat{c}}$, so that overall the VB consists of a union of at most four different flat surfaces depending on the relative orientation and position of the two dimers.

\subsection*{Contact radius}

In order to calculate the excluded volume and surface, $V^*$ and $S^*$, respectively, we require the contact radius $r^*(\mathbf{\hat{r}},\mathbf{\hat{t}})$, which is the value of $r$ for which a dimer $j$ with orientation $\mathbf{\hat{t}}$ and solid angle $\mathbf{\hat{r}}$ is in contact with dimer $i$. Two equal spheres are in contact, when their separation is twice the radius. For two dimers, there are thus four different conditions for contact: $r_{11}=2a$ and likewise for $r_{12}$, $r_{21}$, and $r_{22}$. Solving these four condition for $r$ using the Eqs.~(\ref{dimercontacts}), yields the value of $r$ for contact of sphere $1$ of $i$ and sphere $1$ of $j$, which we denote by $r^*_{ij}$ (Supplementary Fig.~\ref{SFig_dimers}b):
\be
r^*_{11}(\mathbf{\hat{r}},\mathbf{\hat{t}})=\frac{b}{2}\left(\mathbf{\hat{r}}\mathbf{\hat{t}}+\mathbf{\hat{r}}\mathbf{\hat{z}}+\sqrt{(\mathbf{\hat{r}}\mathbf{\hat{t}}+\mathbf{\hat{r}}\mathbf{\hat{z}})^2+\frac{4}{\tilde{\alpha}^2}-2(1+\mathbf{\hat{t}}\mathbf{\hat{z}})}\right).
\ee
Likewise,
\be
r^*_{12}(\mathbf{\hat{r}},\mathbf{\hat{t}})&=&\frac{b}{2}\left(-\mathbf{\hat{r}}\mathbf{\hat{t}}+\mathbf{\hat{r}}\mathbf{\hat{z}}+\sqrt{(\mathbf{\hat{r}}\mathbf{\hat{t}}-\mathbf{\hat{r}}\mathbf{\hat{z}})^2+\frac{4}{\tilde{\alpha}^2}-2(1-\mathbf{\hat{t}}\mathbf{\hat{z}})}\right),\\
r^*_{21}(\mathbf{\hat{r}},\mathbf{\hat{t}})&=&\frac{b}{2}\left(\mathbf{\hat{r}}\mathbf{\hat{t}}-\mathbf{\hat{r}}\mathbf{\hat{z}}+\sqrt{(\mathbf{\hat{r}}\mathbf{\hat{t}}-\mathbf{\hat{r}}\mathbf{\hat{z}})^2+\frac{4}{\tilde{\alpha}^2}-2(1-\mathbf{\hat{t}}\mathbf{\hat{z}})}\right),\\
\label{r22eq}
r^*_{22}(\mathbf{\hat{r}},\mathbf{\hat{t}})&=&\frac{b}{2}\left(-(\mathbf{\hat{r}}\mathbf{\hat{t}}+\mathbf{\hat{r}}\mathbf{\hat{z}})+\sqrt{(\mathbf{\hat{r}}\mathbf{\hat{t}}+\mathbf{\hat{r}}\mathbf{\hat{z}})^2+\frac{4}{\tilde{\alpha}^2}-2(1+\mathbf{\hat{t}}\mathbf{\hat{z}})}\right).\nonumber\\
\ee
The correct overall $r^*$ is then the maximum of the $r^*_{ij}$. This follows simply if we imagine a configuration with fixed relative orientation $\mathbf{\hat{t}}$ and angular position $\mathbf{\hat{r}}$. For a large radius $r$ the two dimers are not in contact. Now decrease $r$. The correct contact radius is then the largest value of $r$ for which the two dimers are in contact for the first time, since for any of the smaller $r^*$ there might be overlap. 

\newpage

\section*{Spherocylinders}

\label{Secsupp_calc_sphero}

A spherocylinder consists of a cylindrical part of length $L$ and radius $a$, with two semi-spheres of radius $a$ as endcaps (Supplementary Fig.~\ref{SFig_spheros}b). This yields the aspect ratio $\alpha=1+L/(2a)$. As for dimers, the hard core boundary of a spherocylinder is parameterized only by the polar angle $\theta_{\rm c}$ due to the rotational symmetry
\begin{widetext}
\be
\label{cs_sc}
c^*(\theta_{\rm c})=a\left\{
\begin{array}{cc}
\tilde{\alpha}\left(\cos(\theta_{\rm c})+\sqrt{\frac{1}{\tilde{\alpha}^{2}}-\sin^2(\theta_{\rm c})}\right),& 0\le\theta_{\rm c}<\arctan(\tilde{\alpha}^{-1})\\ \\
\sin(\theta_{\rm c})^{-1},& \arctan(\tilde{\alpha}^{-1})\le\theta_{\rm c}\le\pi/2,
\end{array}\right.
\ee
\end{widetext}
where $\tilde{\alpha}=\alpha-1=L/(2a)$.

The VB between two spherocylinders is identical to the VB between the line segments at the centre of the cylindrical part. In the following we refer to these line segments as ``rods". As before, we align rod $i$ with the $\mathbf{\hat{z}}$ axis of our coordinate system, so that a point on it is parameterized by the vector $t_i\mathbf{\hat{z}}$ with $t_i \in [-L/2;L/2]$. Likewise, the orientation of rod $j$ is given by $\mathbf{\hat{t}}$, so that a point on rod $j$ is parameterized by $\mathbf{r}+t_j\mathbf{\hat{t}}$, where also $t_j \in [-L/2;L/2]$. 

We solve the two conditions that define the VB as follows. The square of the distance between $s\mathbf{\hat{c}}$ and a point on rod $i$ is
\be
\label{di}
D_i^2 &=& (t_i\mathbf{\hat{z}} - s\mathbf{\hat{c}})^2,
\ee
and likewise the distance between $s\mathbf{\hat{c}}$ and a point on rod $j$
\be
\label{dj}
D_j^2 &=& (\mathbf{r} + t_j\mathbf{\hat{t}} - s\mathbf{\hat{c}})^2.
\ee
Condition 1. then requires:
\be
\frac{\partial D_i^2}{\partial t_i}&=&0,
\label{di0}\\
\frac{\partial D_j^2}{\partial t_j}&=&0.
\label{dj0}
\label{di0dj0}
\ee
This leads to the minimal values
\be
\label{titj1}
t_i^{\rm min}& = &s\mathbf{\hat{c}}\mathbf{\hat{z}} = s (\mathbf{\hat{c}}\mathbf{\hat{z}}),\\
\label{titj2}
t_j^{\rm min} &=& (s\mathbf{\hat{c}}-\mathbf{r}) \mathbf{\hat{t}} = s (\mathbf{\hat{c}}\mathbf{\hat{t}})- r.
\ee
Condition 2. requires:
\be
D^{\rm min}_i=D_j^{\rm min},
\label{didjmin}
\ee
which leads to
\be
\label{mainquad}
(t_i^{\rm min} \mathbf{\hat{z}} - s\mathbf{\hat{c}})^2 = (t_j^{\rm min} \mathbf{\hat{t}} + \mathbf{r} - s\mathbf{\hat{c}})^2.
\ee

Eq.~(\ref{mainquad}) does not take into account that the rods have a finite length $L$, so that $t_i^{min}$ and $t_j^{min}$ are only the correct minimal values when $t_{i}^{\rm min} \in [-L/2,L/2]$ and $t_{j}^{\rm min} \in [-L/2,L/2]$. We refer to this case as a {\it line-line} interaction between the two rods. If $t_i^{min}$ and/or $t_j^{min}$ are not $\in [-L/2,L/2]$ interactions involving the end-points of the rods arise. Overall, one has to distinguish the cases:

\begin{enumerate}
\item{{\it Line-line} interaction: $t_{i}^{\rm min} \in [-L/2,L/2]$
    and $t_{j}^{\rm min} \in [-L/2,L/2]$ (1 case).}
\item{{\it Line-point} interaction between the segment $i$ and an
    end-point of $j$: $t_{i}^{\rm min} \in [-L/2,L/2]$ and $t_j=\pm
    L/2$ (2 cases).}
\item{{\it Point-line} interaction between the segment $j$ and an
    end-point of $i$: $t_{j}^{\rm min} \in [-L/2,L/2]$ and $t_i=\pm
    L/2$ (2 cases).}
\item{{\it Point-point} interaction between the end points of $i$ and
    $j$: $t_i=\pm L/2$ and $t_j=\pm L/2$ (4 cases).}
\end{enumerate}

In the following we use different subscripts in order to refer to the different Voronoi interactions, e.g., $s_{\rm ll}$ for line-line interaction, $s_{lp}$ for a line-point interaction, etc. The separation of the different interactions follows the algorithm outlined in Fig.~\ref{Fig_algorithms}b in the main text. Note that the four point-point interactions are flat surfaces, while interactions involving the line segment are curved.

\subsection*{Line-line interaction}

This case arises if $t_i^{min}$ and $t_j^{min}$ fall inside the length of the segments. The conditions are thus:
\be
\label{tllmin}
t_{i}^{\rm min} &\in& [-L/2,L/2],\qquad t_{j}^{\rm min} \in [-L/2,L/2].
\ee
In this case $t_i^{min}$ and $t_j^{min}$ are given by Eqs.~(\ref{titj1}) and (\ref{titj2}). Substituting these expressions into Eq.~(\ref{mainquad}) then leads to a quadratic equation for the value $s=s_{\rm ll}$ of the boundary:
\be
\label{lleq}
\frac{s_{\rm ll}^2}{r^2} \Big[(\mathbf{\hat{c}}\mathbf{\hat{z}})^2 -(\mathbf{\hat{c}}\mathbf{\hat{t}})^2 \Big]
  + 2 \frac{s_{\rm ll}}{r} \Big[(\mathbf{\hat{c}}\mathbf{\hat{t}})(\mathbf{\hat{r}}\mathbf{\hat{t}}) - \mathbf{\hat{r}}\mathbf{\hat{c}} \Big]+ 1 - (\mathbf{\hat{r}}\mathbf{\hat{t}})^2 = 0.
\ee
The correct solution of this equation is the real and positive one. Clearly, the line-line Voronoi boundary between the two rods scales with the separation $r$.

Eqs.~(\ref{tllmin}) are satisfied when
\be
-L/2\le s_{\rm ll}\mathbf{\hat{c}}\mathbf{\hat{z}}\le L/2,\quad {\rm and}\quad -L/2\le(s_{\rm ll}\mathbf{\hat{c}}-\mathbf{r})\mathbf{\hat{t}}\le L/2,
\ee
which defines the separation lines for the line interactions on each of the two spherocylinders in Fig.~\ref{Fig_algorithms}b. The VB due to the line-line interaction is illustrated further in the Supplementary Fig.~\ref{SFig_spheros}a: A sphere centred at the VB touches both rods $i$ and $j$ for a unique radius.

\subsection*{Line-point interaction}

In this case $t_i^{\rm min}$ falls along the line segment $i$ and $t_j^{\rm min}$ is at one of the end points of rod $j$. We choose the top of $\mathbf{t}_j$ as the point, indicated by a subscript $1$ and we obtain:
\be
t_{i}^{\rm min} &\in& [-L/2,L/2],\qquad t_{j}^{\rm min} = L/2.
\ee
Substituting the Eq.~(\ref{titj1}) for $t_i^{min}$ and $t_{j}^{\rm min} = L/2$ into Eq.~(\ref{mainquad}) then leads to a quadratic equation for $s=s_{\rm lp_1}$, where the index $p_1$ refers to the top point:
\be
\label{lpeq}
  \frac{s_{\rm lp_1}^2}{r^2} (\mathbf{\hat{c}}\mathbf{\hat{z}})^2 -2\frac{s_{\rm lp_1}}{r} \Big[ (\mathbf{\hat{r}}\mathbf{\hat{c}}) 
  + \frac{L}{2r} (\mathbf{\hat{c}}\mathbf{\hat{t}}) \Big ] + \left(\frac{L}{2r}\right)^2 + \frac{L}{r} (\mathbf{\hat{r}}\mathbf{\hat{t}}) + 1=0. 
\ee
The corresponding expression for the Voronoi boundary with respect to the bottom point $s_{\rm lp_2}$, where $t_{j}^{\rm min} = -L/2$, simply follows by setting $L\to -L$ in Eq.~(\ref{lpeq}). The conditions for the two line-point interactions are then
\be
-L/2\le s_{\rm lp_1}\mathbf{\hat{c}}\mathbf{\hat{z}}\le L/2\quad &{\rm and}&\quad (s_{\rm lp_1}\mathbf{\hat{c}}-\mathbf{r})\mathbf{\hat{t}}\ge L/2\\
-L/2 \le s_{\rm lp_2}\mathbf{\hat{c}}\mathbf{\hat{z}}\le L/2\quad &{\rm and}&\quad (s_{\rm lp_2}\mathbf{\hat{c}}-\mathbf{r})\mathbf{\hat{t}}\le -L/2.
\ee

\subsection*{Point-line interaction}

This interaction is analogous to line-point. The conditions are:
\be
t_{i}^{\rm min} &=& L/2,\qquad t_{j}^{\rm min} \in[-L/2,L/2].
\ee
Substituting $t_{i}^{\rm min} = L/2$ for the top point and Eq.~(\ref{titj2}) into Eq.~(\ref{mainquad}) leads to
\be
\label{pleq}
&&\frac{s_{\rm p_1l}^2}{r^2} (\mathbf{\hat{c}}\mathbf{\hat{t}})^2 + 2\frac{s_{\rm p_1l}}{r} [(\mathbf{\hat{r}}\mathbf{\hat{c}}) - (\mathbf{\hat{c}}\mathbf{\hat{t}}) (\mathbf{\hat{r}}\mathbf{\hat{t}})]\nonumber\\
&&- \frac{s_{\rm p_1l}}{r}\frac{L}{r} (\mathbf{\hat{c}}\mathbf{\hat{z}})+ \left(\frac{L}{2r}\right)^2 + (\mathbf{\hat{r}}\mathbf{\hat{t}})^2- 1 = 0. 
\ee
Likewise for $s_{\rm p_2l}$. The conditions for the two point-line interactions are then
\be
s_{\rm p_1l}\mathbf{\hat{c}}\mathbf{\hat{z}}\ge L/2,\quad &{\rm and}&\quad -L/2\le (s_{\rm p_1l}\mathbf{\hat{c}}-\mathbf{r})\mathbf{\hat{t}}\le L/2\\
s_{\rm p_2l}\mathbf{\hat{c}}\mathbf{\hat{z}}\le -L/2,\quad &{\rm and}&\quad -L/2\le (s_{\rm p_2l}\mathbf{\hat{c}}-\mathbf{r})\mathbf{\hat{t}}\le L/2.
\ee

\subsection*{Point-point interaction}

In this case the two points $t_i^{\rm min}$ and $t_j^{\rm min}$ are both fixed and equal to $L/2$ or $-L/2$. Writing
\be
\label{tt}
t_i^{\rm min} = L_i/2,\qquad t_j^{\rm min} = L_j/2,
\ee
where $L_i=\pm L$ and $L_j=\pm L$ for the top and bottom points on each of the rods, we find for the solution of Eq.~(\ref{mainquad}) with Eqs.~(\ref{tt}):
\be
\label{ppsol}
  s_{pp} = r\,\frac{1 + \frac{L_j}{r} (\mathbf{\hat{r}}\mathbf{\hat{t}})}{2(\mathbf{\hat{r}}\mathbf{\hat{c}}) + \frac{L_j}{r} 
(\mathbf{\hat{c}}\mathbf{\hat{t}}) - \frac{L_i}{r}(\mathbf{\hat{c}}\mathbf{\hat{z}})}.
\ee
Here, the interactions for, e.g., the two top points $s_{\rm p_1p_1}$ is obtained by setting $L_i=L_j=L$. Likewise for the other point interactions. The conditions for the four different point-point Voronoi boundaries are then
\be
s_{\rm p_1p_1}\mathbf{\hat{c}}\mathbf{\hat{z}}\ge L/2,\quad &{\rm and}&\quad (s_{\rm p_1p_1}\mathbf{\hat{c}}-\mathbf{r})\mathbf{\hat{t}}\ge L/2,\\
s_{\rm p_1p_2}\mathbf{\hat{c}}\mathbf{\hat{z}}\ge L/2,\quad &{\rm and}&\quad (s_{\rm p_1p_2}\mathbf{\hat{c}}-\mathbf{r})\mathbf{\hat{t}}\le -L/2,\\
s_{\rm p_2p_1}\mathbf{\hat{c}}\mathbf{\hat{z}}\le -L/2,\quad &{\rm and}&\quad (s_{\rm p_2p_1}\mathbf{\hat{c}}-\mathbf{r})\mathbf{\hat{t}}\ge L/2,\\
s_{\rm p_2p_2}\mathbf{\hat{c}}\mathbf{\hat{z}}\le -L/2,\quad &{\rm and}&\quad (s_{\rm p_2p_2}\mathbf{\hat{c}}-\mathbf{r})\mathbf{\hat{t}}\le -L/2.
\ee

In the limit $L/r\rightarrow 0$, we recover from Eq.~(\ref{ppsol}) the Voronoi boundary between two equal spheres, Eq.~(\ref{VBsphere}).

\subsection*{Contact radius}
 
In order to determine the contact radius $r^*(\mathbf{\hat{r}},\mathbf{\hat{t}})$ of two spherocylinders, one has to distinguish the possible contacts of the spherical endcaps and of the cylindrical segments. As before, we denote a point on rod $i$ by $t_i\mathbf{\hat{z}}$ and a point on rod $j$ by $\mathbf{r}+t_j\mathbf{\hat{t}}$. The squared distance between these two points is
 \be
 \label{Dcontact}
 D^2(\mathbf{r},\mathbf{\hat{t}},t_i,t_j)&=&(t_i\mathbf{\hat{z}}-(\mathbf{r}+t_j\mathbf{\hat{t}}))^2\nonumber\\
 &=&t_i^2+t_j^2+r^2+2r(t_j(\mathbf{\hat{r}}\mathbf{\hat{t}})-t_i(\mathbf{\hat{r}}\mathbf{\hat{z}}))-2t_it_j(\mathbf{\hat{z}}\mathbf{\hat{t}}).\nonumber\\
\ee
The two spherocylinders are in contact when the minimum of $D^2$ with respect to $t_i$ and $t_j$, i.e., the minimal squared separation, is the square of the diameter $(2a)^2$. Solving $\partial D^2/\partial t_i=0$ and $\partial D^2/\partial t_j=0$ yields the two minimal positions
\be
t^*_i&=&r\frac{(\mathbf{\hat{r}}\mathbf{\hat{z}})-(\mathbf{\hat{r}}\mathbf{\hat{t}})(\mathbf{\hat{z}}\mathbf{\hat{t}})}{1-(\mathbf{\hat{z}}\mathbf{\hat{t}})^2}= r A_i\\
t^*_j&=&r\frac{(\mathbf{\hat{r}}\mathbf{\hat{z}})(\mathbf{\hat{z}}\mathbf{\hat{t}})-(\mathbf{\hat{r}}\mathbf{\hat{t}})}{1-(\mathbf{\hat{z}}\mathbf{\hat{t}})^2}= r A_j,
\ee
which define $A_i$ and $A_j$. Substituting these expressions into Eq.~(\ref{Dcontact}) and solving for $r$ under the condition $D^2=4a^2$ yields the contact radius
\be
r_{\rm ll}^*(\mathbf{\hat{r}},\mathbf{\hat{t}})= \frac{2 a}{\sqrt{1+(A_i\mathbf{\hat{z}}-A_j\mathbf{\hat{t}})^2+2(A_j(\mathbf{\hat{r}}\mathbf{\hat{t}})-A_i(\mathbf{\hat{r}}\mathbf{\hat{z}}))}}.
\ee
This contact radius does not take into account the finite length of the spherocylinders and is only valid for $t^*_i\in[-L/2,L/2]$ and $t^*_j\in[-L/2,L/2]$. In fact, $r_{\rm ll}^*$ is the contact between the line segments (indicated by the subscript as before). As for the different Voronoi interactions one has to distinguish further the line-point, point-line and line-line contacts in addition to the line-line one (Supplementary Fig.~\ref{SFig_spheros}b).

For the line-point contact one has to consider $t_j=\pm L/2$, so one has to solve
\be
\frac{\partial}{\partial t_i}D^2\left(\mathbf{r},\mathbf{\hat{t}},t_i,\pm \frac{L}{2}\right)=0
\ee
to find the minimal $t^{\rm *lp}_i$. Substituting this value back into $D^2$ and solving $D^2=4a^2$ for $r$ yields the two line-point contact radii, which are valid when $t_i^{\rm *lp}\in[-L/2,L/2]$. For the point-line contact one has to consider $t_i=\pm L/2$, so that the corresponding equation is given by
\be
\frac{\partial}{\partial t_j}D^2\left(\mathbf{r},\mathbf{\hat{t}},\pm\frac{L}{2},t_j\right)=0
\ee
determines the minimal $t_j^{\rm *pl}$. Substituting this value back into $D^2$ and solving $D^2=4a^2$ for $r$ yields the two point-line contact radii. These are valid when $t_j^{\rm *pl}\in[-L/2,L/2]$. For the point-point contact one can solve directly
\be
D^2\left(\mathbf{r},\mathbf{\hat{t}},\pm\frac{L}{2},\pm\frac{L}{2}\right)=4 a^2
\ee
 for $r$, which yields four different point-point contact radii.

Overall, one thus obtains 9 possible different valid values for the contact radius $r^*(\mathbf{\hat{r}},\mathbf{\hat{t}})$, similar to the different Voronoi interactions. The unique correct radius is then the maximum of all positive and real ones.

\section*{Calculation of the packing fraction}

Here, we summarize our method to calculate the packing fraction of dimers and spherocylinders, shown in Fig.~\ref{Fig_results}b in the main text. We first calculate $V^*$ and $S^*$ numerically for a range of $\mathbf{c}$ values. The excluded volume is defined as $V^*=\left<\Omega-\Omega\cap V_{\rm ex}\right>_{\mathbf{\hat{t}}}$, which can be expressed as an orientational average over a volume integral:
\be
\label{exvol}
V^*(\mathbf{c})&=&\left<\int \D \mathbf{r}\,\Theta(r-r^*(\mathbf{\hat{r}},\mathbf{\hat{t}}))\Theta(c-s(\mathbf{r},\mathbf{\hat{t}},\mathbf{\hat{c}}))\Theta(s(\mathbf{r},\mathbf{\hat{t}},\mathbf{\hat{c}}))\right>_{\mathbf{\hat{t}}}.\nonumber\\
\ee
We parametrize these integrals in spherical coordinates and denote with $\theta_r$, the polar angle of the position and with $\beta_r$ the azimuthal angle of the position. The corresponding orientational angles have a subscript $t$. Eq.~(\ref{exvol}) can then be written in terms of the multi-dimensional integral
\begin{widetext}
\be
\label{vstar_calc}
V^*(c,\theta_{\rm c})&=&\frac{1}{2\pi}\int_0^{\pi}\D \theta_r\int_{-\pi}^\pi\D\beta_r\int_0^{\pi/2}\D \theta_t\int_{-\pi}^\pi\D\beta_t\int_{r^*(\theta_r,\beta_r,\theta_t,\beta_t)}^\infty\D r\,r^2 \sin(\theta_t)\sin(\theta_r)\Theta[c-s(r,\theta_r,\beta_r,\theta_t,\beta_t,\theta_{\rm c})]\Theta[s(r,\theta_r,\beta_r,\theta_t,\beta_t,\theta_{\rm c})].\nonumber\\
\ee
\end{widetext}
Here, the integration limits of the $\theta_t$ integration only take distinct orientations into account. Eq.~(\ref{vstar_calc}) is a five dimensional integral, which we calculate numerically using a Monte-Carlo method for a given $\mathbf{c}$.

The excluded surface is defined $S^*=\left<\partial V_{\rm ex}\cap\Omega\right>_{\mathbf{\hat{t}}}$, which can be expressed as an orientational average over a surface integral:
\be 
\label{exsur}
S^*(\mathbf{c})&=&\left<\left.\oint \D \mathbf{\hat{r}}\,\Theta(c-s(\mathbf{r},\mathbf{\hat{t}},\mathbf{\hat{c}}))\Theta(s(\mathbf{r},\mathbf{\hat{t}},\mathbf{\hat{c}}))\right|_{r=r^*(\mathbf{\hat{r}},\mathbf{\hat{t}})}\right>_{\mathbf{\hat{t}}},
\ee
Here, one has to take into account the surface element for a non-constant radius $r^*(\mathbf{\hat{r}},\mathbf{\hat{t}})$. Using the same parametrization as for the excluded volume, the surface element can be calculated and yields
\be
\D \mathbf{\hat{r}}=r^*\sqrt{\left(r^{*2}+\left(\frac{\partial r^*}{\partial \theta_r}\right)^2\right)\sin^2(\theta_r)+\left(\frac{\partial r^*}{\partial \beta_r}\right)^2}\D\theta_r\D\beta_r,
\ee
which recovers the usual surface element $\D \mathbf{\hat{r}}=r^{*2}\sin(\theta_r)\D\theta_r\D\beta_r$ for $r^*={\rm const}$. Eq.~(\ref{exsur}) can thus be written in terms of the multi-dimensional integral
\begin{widetext}
\be
\label{sstar_calc}
S^*(c,\theta_{\rm c})&=&\frac{1}{2\pi}\int_0^{\pi}\D \theta_r\int_{-\pi}^\pi\D\beta_r\int_0^{\pi/2}\D \theta_t\int_{-\pi}^\pi\D\beta_t \sin(\theta_t)r^*\sqrt{\left(r^{*2}+\left(\frac{\partial r^*}{\partial \theta_r}\right)^2\right)\sin^2(\theta_r)+\left(\frac{\partial r^*}{\partial \beta_r}\right)^2}\nonumber\\
&&\times\Theta[c-s(r^*,\theta_r,\beta_r,\theta_t,\beta_t,\theta_{\rm c})]\Theta[s(r^*,\theta_r,\beta_r,\theta_t,\beta_t,\theta_{\rm c})], 
\ee
\end{widetext}
where $r^*=r^*(\theta_r,\beta_r,\theta_t,\beta_t)$. Eq.~(\ref{sstar_calc}) can also be computed numerically using Monte-Carlo for a given $\mathbf{c}$.

In the next step we determine the surface density $\sigma(z)$ with the method outlined in the section Methods: We generate local configurations of $z$ contacting particles and determine the probability density function $p_{\rm m}(c_{\rm m},\mathbf{\hat{c}})$ of the minimal VB along a direction $\mathbf{\hat{c}}$. This yields the average
\be
\left<S^*(c_{\rm m},\mathbf{\hat{c}})\right>=\int_{c^*}^\infty S^*(y,\mathbf{\hat{c}})p_{\rm m}(y,\mathbf{\hat{c}})\D y,
\ee
and the surface density follows via Eq.~(\ref{dens2}) for integer values of $z$
\[
\sigma(z)=\frac{1}{\left<\left<S^*(c_{\rm m},\mathbf{\hat{c}})\right>\right>_{\mathbf{\hat{c}}}}.
\]

The average Voronoi volume can then be calculated by solving the
self-consistent equation~(\ref{integral}) numerically for a given integer
$z$. The volume integral on the right hand side of Eq.~(\ref{integral}) with Eq.~(\ref{cdf})
reads explicitly for the rotationally symmetric dimers and
spherocylinders \be
\overline{W}(z)&=&V_\alpha+4\pi\int_0^{\pi/2}\D\theta_{c}\sin(\theta_{\rm c})\int_{c^*(\theta_{\rm c})}^\infty\D
c\,c^2\exp\left\{-\frac{V^*(c,\theta_{\rm c})}{\overline{W}(z)-V_\alpha}-\sigma(z)\,S^*(c,\theta_{\rm c})\right\}.
\ee In order to solve this equation numerically we calculate the
two-dimensional integral on the right hand side using our numerically
obtained $V^*$, $S^*$, and $\sigma(z)$ for a given $z$ over a range of
$\overline{W}=x$ values. This yields a function $G(x)$. The average
Voronoi volume $\overline{W}$ is then the value of $x$ that satisfies $G(x)=x$ and
the packing fraction $\phi(z,\alpha)$ follows as $V_\alpha/\overline{W}$. For
fractional $z$ that are predicted from our evaluation of degenerate
configurations (Methods), we use a linear
interpolation to obtain $\phi(z(\alpha),\alpha)$.

\section*{Analytic continuation of the spherical random close packing}

\label{Secsupp_ac}

Close to the spherical point, the self-consistent Eq.~(\ref{integral}) can be solved analytically and allows the calculation of an analytic continuation from the RCP point. The key is to introduce suitable approximations of $V^*$ and $S^*$ for $\alpha$ close to $1$. We assume that, as the particles are deformed from the sphere, the change in the excluded volume and surface terms is dominated by the hard-core exclusion, while the change due to the Voronoi interaction can be neglected. This means that $V^*$ and $S^*$ are given by the spherical excluded volume and surface, but shifted by $c^*(\mathbf{\hat{c}})-a$:
\be
V^*(\mathbf{c})&=&V^*_1(c-(c^*(\mathbf{\hat{c}})-a)),\\
\label{sstar_ac}
S^*(\mathbf{c})&=&S^*_1(c-(c^*(\mathbf{\hat{c}})-a)).
\ee
Here, $V^*_1$ and $S^*_1$ are the corresponding expressions for spheres \cite{Song08}:
\be
\label{spherical_ex1}
V_1^*(c)&=&V_1\left(\left(\frac{c}{a}\right)^3-4+3\frac{a}{c}\right),\\
\label{spherical_ex2}
S_1^*(c)&=&2S_1\left(1-\frac{a}{c}\right),
\ee
with $V_1$ and $S_1$ denoting the volume and surface of a sphere with radius $a$. In the following, the subscript $1$ always refers to quantities in spherical packings with $\alpha=1$. With these approximations, the self-consistent Eq.~(\ref{integral}) becomes
\be
\label{selfac}
\overline{W}&=&V_\alpha+\oint\D\mathbf{\hat{c}}\int_{c^*(\mathbf{\hat{c}})}^\infty \D c\, c^2\exp\left\{-\frac{1}{\overline{W}-V_\alpha}V^*_1(c-(c^*(\mathbf{\hat{c}})-a))-\sigma(z)S^*_1(c-(c^*(\mathbf{\hat{c}})-a))\right\}.
\ee
We transform the integration variable into
\be
x=\frac{c-(c^*(\mathbf{\hat{c}})-a)}{a}.
\ee
Substituting into Eq.~(\ref{selfac}) the expressions for $V^*_1$ and $S^*_1$, Eqs.~(\ref{spherical_ex1}) and (\ref{spherical_ex2}), and dividing the equation by the sphere volume $V_1$ leads to
\be
\omega&=&\frac{3}{4\pi}\oint\D\mathbf{\hat{c}}\int_{1}^\infty \D x\, \left(x+\frac{c^*(\mathbf{\hat{c}})}{a}-1\right)^2\exp\left\{-\frac{1}{\omega}\left(x^3+\frac{3}{x}-4\right)-\tilde{\sigma}(z)\left(1-\frac{1}{x}\right)\right\},
\ee
where we define the quantities
\be
\label{freevol_def}
\omega=\frac{\overline{W}-V_\alpha}{V_1},\qquad \tilde{\sigma}(z)=2S_1\sigma(z).
\ee
Rearranging terms yields
\be
\label{self2}
\omega&=&3\left<\int_{1}^\infty \D x\, \left(x+\frac{c^*(\mathbf{\hat{c}})}{a}-1\right)^2\exp\left\{-\frac{1}{\omega}\left(x^3+\left(3-\tilde{\sigma}(z)\omega\right)\frac{1}{x}-4-\tilde{\sigma}(z)\omega\right)\right\}\right>_{\mathbf{\hat{c}}}.
\ee
Now we use the identity
\begin{widetext}
\be
-\omega\frac{\D}{\D x}\exp\left\{-\frac{1}{\omega}\left(x^3+\left(3-\tilde{\sigma}(z)\omega\right)\frac{1}{x}-4-\tilde{\sigma}(z)\omega\right)\right\}&=&\left(3x^2-\left(3-\tilde{\sigma}(z)\omega\right)\frac{1}{x^2}\right)\exp\left\{-\frac{1}{\omega}\left(x^3+\left(3-\tilde{\sigma}(z)\omega\right)\frac{1}{x}-4-\tilde{\sigma}(z)\omega\right)\right\}\nonumber\\
\ee
to obtain from Eq.~(\ref{self2})
\be
\label{selfac2}
0&=&\left<\int_{1}^\infty \D x\, \left(\left(3-\tilde{\sigma}(z)\omega\right)\frac{1}{x^2}+6x(c^*(\mathbf{\hat{c}})/a-1)+3(c^*(\mathbf{\hat{c}})/a-1)^2\right)\exp\left\{-\frac{1}{\omega}\left(x^3+\left(3-\tilde{\sigma}(z)\omega\right)\frac{1}{x}-4-\tilde{\sigma}(z)\omega\right)\right\}\right>_{\mathbf{\hat{c}}}.
\ee
\end{widetext}
In the spherical limit $\alpha\to 1$, we have $c^*\to a$ and one can show that \cite{Song08}
\be
\label{sphere_sigma}
\tilde{\sigma}_1(z)=z\sqrt{3}/2.
\ee
In this case Eq.~(\ref{selfac2}) becomes
\be
0&=&\left(3-\tilde{\sigma}_1(z)\omega_1\right)\int_{1}^\infty \D x\, \frac{1}{x^2}\exp\left[-\frac{1}{\omega}_1\left(x^3+\left(3-\tilde{\sigma}_1(z)\omega_1\right)\frac{1}{x}-4-\tilde{\sigma}_1(z)\omega_1\right)\right],\nonumber\\
\ee
which has the exact solution
\be
3-\tilde{\sigma}_1(z)\omega_1=0,
\ee
so that the free volume becomes
\be
\label{sphere_freevol}
\omega_1(z)=\frac{3}{\tilde{\sigma}_1(z)}=\frac{2\sqrt{3}}{z},
\ee
using Eq.~(\ref{sphere_sigma}). In order to solve Eq.~(\ref{selfac2}) for $\alpha \neq 1$, we approximate
\be
e^{-\left(3-\tilde{\sigma}(z)\omega\right)\frac{1}{\omega\, x}}\approx 1-\left(3-\tilde{\sigma}(z)\omega\right)\frac{1}{\omega\,x},
\ee
which is an appropriate approximation since the dominant term in the exponent for the given integration limits is $x^3$ and $\tilde{\sigma}(z)$ is of order $1$ for small aspect ratios. This leads to
\be
0&=&\left<\int_{1}^\infty \D x\, \left(\left(3-\tilde{\sigma}(z)\omega\right)\frac{1}{x^2}+6x(c^*(\mathbf{\hat{c}})/a-1)+3(c^*(\mathbf{\hat{c}})/a-1)^2\right)\left(1-\left(3-\tilde{\sigma}(z)\omega\right)\frac{1}{\omega\,x}\right)e^{-x^3/\omega}\right>_{\mathbf{\hat{c}}},
\ee
so that the integration over $x$ and the orientational average become independent. We obtain further
\be
0&=&\int_{1}^\infty \D x\, \left(\left(3-\tilde{\sigma}(z)\omega\right)\frac{1}{x^2}+6x\left<(c^*(\mathbf{\hat{c}})/a-1)\right>_{\mathbf{\hat{c}}}+3\left<(c^*(\mathbf{\hat{c}})/a-1)^2\right>_{\mathbf{\hat{c}}}\right)\left(\omega-\left(3-\tilde{\sigma}(z)\omega\right)\frac{1}{x}\right)e^{-x^3/\omega},
\ee
or, after rewriting the integrals,
\begin{widetext}
\be
\label{selfapprox}
0&=&-\left(3-\tilde{\sigma}(z)\omega\right)^2f_{-3}(\omega)-\left(3-\tilde{\sigma}(z)\omega\right)\left[6\left<(c^*(\mathbf{\hat{c}})/a-1)\right>_{\mathbf{\hat{c}}}f_0(\omega)+3\left<(c^*(\mathbf{\hat{c}})/a-1)^2\right>_{\mathbf{\hat{c}}}f_{-1}(\omega)-\omega f_{-2}(\omega)\right]\nonumber\\
&&+6\omega\left<(c^*(\mathbf{\hat{c}})/a-1)\right>_{\mathbf{\hat{c}}}f_1(\omega)+3\omega\left<(c^*(\mathbf{\hat{c}})/a-1)^2\right>_{\mathbf{\hat{c}}}f_0(\omega).
\ee
\end{widetext}
This equation is quadratic in $3-\tilde{\sigma}(z)\omega$ and contains the basic integrals
\be
f_n(y)=\int_1^\infty\D x\,x^n\,e^{-x^3/y},
\ee
which can not be expressed in closed form. The solution of Eq.~(\ref{selfapprox}) is
\be
\label{selfapprox2}
3-\tilde{\sigma}(z)\omega=F_\alpha(\omega),
\ee
where we indicate the dependence on $\alpha$ explicitly. In the spherical limit, we have $F_1(\omega)=0$ and we recover the spherical result. By expanding the function $F_\alpha(\omega)$ we therefore obtain an analytical continuation of the spherical solution. In the following we neglect quadratic terms in the deviation from the sphere. Expanding $F_\alpha(\omega)$ into a Taylor series up to linear orders in $\tilde{\alpha}=\alpha-1$ leads to
\be
\label{selfapprox3}
3-\tilde{\sigma}(z)\omega=-6 M_{\rm b} h(\omega)\tilde{\alpha},
\ee
where
\be
\label{hfunc}
h(y)=\frac{f_{1}(y)}{f_{-2}(y)},
\ee
and the constant $M_{\rm b}$ denotes the relative first-order deviation of the object boundary from the sphere (the subscript b refers to ``boundary"):
\be
\label{Mb}
M_{\rm b}=\frac{1}{a}\left.\frac{\D}{\D\alpha}\left<c^*(\mathbf{\hat{c}})\right>_{\mathbf{\hat{c}}}\right|_{\alpha=1}.
\ee

We are interested in an analytic continuation of the spherical RCP point as the sphere is deformed. At RCP the coordination number is given by the isostatic value $\bar{z}=6$, so that the free volume Eq.~(\ref{sphere_freevol}) at RCP becomes $\bar{\omega}_1=\omega_1(\bar{z})=1/\sqrt{3}$. If we expand $h(\omega)$ around $\bar{\omega}_1$ to linear orders in $\tilde{\alpha}$ we obtain from Eq.~(\ref{selfapprox3})
\be
3-\tilde{\sigma}(z)\omega=-6M_{\rm b}(h(\bar{\omega}_1)+h'(\bar{\omega}_1))(\omega-\bar{\omega}_1)\tilde{\alpha},
\ee
which can be solved for $\omega$
\be
\label{omegasol}
\omega=\frac{3+6M_{\rm b}(h(\bar{\omega}_1)-h'(\bar{\omega}_1)\bar{\omega}_1)\tilde{\alpha}}{\tilde{\sigma}(z)-6 M_{\rm b}h'(\bar{\omega}_1)\tilde{\alpha}}.
\ee
By factoring out the spherical surface density at RCP, $\tilde{\sigma}_1(\bar{z})$, in the denominator and using $\bar{\omega}_1=3/\tilde{\sigma}_1(\bar{z})$ from Eq.~(\ref{sphere_freevol}) we obtain further
\be
\label{omegasol2}
\omega=\bar{\omega}_1\frac{1+2M_{\rm b}(h(\bar{\omega}_1)-h'(\bar{\omega}_1)\bar{\omega}_1)\tilde{\alpha}}{\tilde{\sigma}(z)/\tilde{\sigma}_1(\bar{z})-2 M_{\rm b}h'(\bar{\omega}_1)\bar{\omega}_1\tilde{\alpha}}.
\ee
For simplicity in the notation, we introduce the two functions
\be
g_1(y)&=&2(h(y)-h'(y)y)\\
g_2(y)&=&2h'(y)y.
\ee
We also multiply $\omega$ by $V_1/V_\alpha$, which yields the reduced free volume per particle: $\omega_\alpha=\omega V_1/V_\alpha$. In turn, $\omega_\alpha$ is directly related to the packing fraction due to Eq.~(\ref{freevol_def})
\be
\phi=\frac{1}{1+\omega_\alpha}.
\ee
With Eq.~(\ref{omegasol2}) we obtain for $\omega_\alpha$
\be
\label{omegasol3}
\omega_\alpha=\bar{\omega}_1\frac{1+M_{\rm b}g_1(\bar{\omega}_1)\tilde{\alpha}}{\tilde{\sigma}(z)/\tilde{\sigma}_1(\bar{z})-M_{\rm b}g_2(\bar{\omega}_1)\tilde{\alpha}}\frac{V_1}{V_\alpha}.
\ee
The crucial step is then to find a suitable approximation for the surface density close to the spherical point. For spheres, the density is linear in $z$, Eq.~(\ref{sphere_sigma}). Since $z$ increases rapidly from the spherical point \cite{Donev07}, we assume that the increase in the surface density is dominated by the increase in the coordination number. Consequently,
\be
\label{sigmaapprox}
\frac{\tilde{\sigma}(z)}{\tilde{\sigma}_1(\bar{z})}\approx\frac{z(\alpha)}{\bar{z}}\approx 1+M_z\tilde{\alpha}.
\ee
In the last step, we have introduced the first-order deviation of the coordination number from the isostatic value
\be
M_z=\frac{1}{\bar{z}}\left.\frac{\D}{\D \alpha}z(\alpha)\right|_{\alpha=1}.
\ee
Substituting Eq.~(\ref{sigmaapprox}) into Eq.~(\ref{omegasol3}) leads to our final result for the reduced free volume per particle
\be
\label{omegasol4}
\omega_\alpha=\bar{\omega}_1\frac{1+M_{\rm b}g_1(\bar{\omega}_1)\tilde{\alpha}}{[1+(M_z-M_{\rm b}g_2(\bar{\omega}_1))\tilde{\alpha}][1+M_{\rm v}\tilde{\alpha}]},
\ee
where we use the first-order variation of the object volume
\be
\label{Mv}
M_{\rm v}=\frac{1}{V_1}\left.\frac{\D}{\D \alpha}V_\alpha\right|_{\alpha=1}=\frac{1}{V_1}\left.\frac{\D}{\D \alpha}\left<c^*(\mathbf{\hat{c}})^3\right>_{\mathbf{\hat{c}}}\right|_{\alpha=1}.
\ee

By expressing $\tilde{\alpha}$ in terms of $z$ using Eq.~(\ref{sigmaapprox}) one can also derive an exact expression for $\phi(z)$, namely Eq.~(\ref{continuation}) in the main text (with $\bar{\omega}_1\to \omega_1$ for simplicity in the notation). At the isostatic value $z=\bar{z}$, Eq.~(\ref{continuation}) recovers the spherical RCP value $\phi(\bar{z})=(1+\bar{\omega}_1)^{-1}$. The inversion of Eq.~(\ref{continuation}) can be performed exactly by solving a quadratic equation for $z(\phi)$, leading to the analytic continuation of the spherical equation of state. For a considerable range of $\phi$ values, the resulting $z(\phi)$ curves are in excellent agreement with the solution obtained by numerically integrating the exact $V^*$ and $S^*$ for dimers and spherocylinders, as shown in Fig.~\ref{Fig_results}c in the main text. Moreover, the maximal packing densities of dimers and spherocylinders from simulations lie very close to the predicted $z(\phi)$ continuation.

Note that Eq.~(\ref{omegasol4}) will lead to different results for the
continuation depending on the boundary parametrization $c^*(\mathbf{\hat{c}})$
used for the particular shape. For example, the parametrization
Eq.~(\ref{cs_sc}) for spherocylinders implies a linearly increasing
object volume with $\tilde{\alpha}$:
$V_\alpha=V_1(1+1.5\tilde{\alpha})$. Instead, one could use a
parametrization that leaves the volume constant $V_\alpha=V_1$, by
rescaling the radius $a$ in Eq.~(\ref{cs_sc}) by the factor
$(1+1.5\tilde{\alpha})^{1/3}$, resulting in a different
$\omega_\alpha$ for the {\it same} aspect ratio. This is not a
physical inconsistency of the theory, but originates in the
approximations for $V^*$ and $S^*$ given by Eq.~(\ref{sstar_ac}),
which are proportional to $V_1$ and $S_1$ and thus also depend on
$a$. Rescaling only the radius of the spherocylinder, while leaving
$V_1$ and $S_1$ unchanged, therefore gives rise to different
approximations. In our approximation, the radii of $V_1$ and $S_1$ are
both identical to the radius $a$ of the spherical components of the
dimers and spherocylinders for all aspect ratios, and thus $V^*$ and
$S^*$ are locally given by the spherical excluded volume and surface.

Supplementary Table~\ref{suptable} summarizes the values of $M_z$, $M_{\rm b}$, and $M_{\rm v}$
for the rotationally symmetric shapes dimers, spherocylinders, and
prolate/oblate ellipsoids. The values of the remaining constants in
Eq.~(\ref{omegasol4}) are: \be \bar{\omega}_1=1/\sqrt{3},\qquad
g_1(\bar{\omega}_1)=2.177,\qquad g_2(\bar{\omega}_1)=0.615.  \ee The
resulting analytic continuations are plotted in the inset of
Fig.~\ref{Fig_results}c in the main text.

From Eq.~(\ref{continuation}) we derive a simple condition such that a
given shape increases the packing density beyond RCP upon deformation. The condition $\phi'(\bar{z})>0$
leads to the inequality \be
\label{condition}
\left[g_1(\bar{\omega}_1)+g_2(\bar{\omega}_1)\right]\,\frac{M_{\rm b}}{M_z}-\frac{M_{\rm v}}{M_z}<1.
\ee 
For prolate shapes we have $M_z\ge 0$, so that
Eq.~(\ref{condition}) is already satisfied if
$\left[g_1(\omega_1)+g_2(\omega_1)\right]M_{\rm b}-M_{\rm v}<0$, which is valid
for dimers, spherocylinders, and prolate ellipsoids. A similar
argument holds for oblate shapes, where $M_z\le 0$.

\end{document}